\documentclass[letterpaper, 10 pt, conference]{ieeeconf}  

\IEEEoverridecommandlockouts
\usepackage{cite}
\usepackage{amsmath,amssymb,amsfonts}
\usepackage{subcaption} 
\usepackage[most]{tcolorbox}

\usepackage{float}

\usepackage{lmodern}
\usepackage{booktabs}
\usepackage{adjustbox}

\usepackage[top=20.1mm,bottom=19.3mm,left=18.9mm,right=18.9mm]{geometry}

\usepackage{wrapfig}
\usepackage{lipsum}
\usepackage{amssymb}
\usepackage{amsmath}
\usepackage{siunitx}
\usepackage{mathtools}
\usepackage[hidelinks]{hyperref}
\usepackage{multirow}

\let\labelindent\relax
\usepackage[inline]{enumitem}
\usepackage[ruled,vlined,linesnumbered,noend]{algorithm2e}
\usepackage[noend]{algpseudocode}
\usepackage{ifthen}

\usepackage{dsfont}
\usepackage{balance}
\usepackage{dirtytalk}
\usepackage{afterpage}
\usepackage{bm}

\usepackage{times}

\allowdisplaybreaks

\newcommand{\diag}{\text{diag}}

\newcommand{\G}{\mathcal{G}}

\newcommand{\N}{\mathbb{N}}

\newcommand{\ra}{\rightarrow}
\newcommand{\R}{\mathbb{R}}

\newtheorem{remark}{Remark}
\newtheorem{assumption}{Assumption}
\newtheorem{proposition}{Proposition}
\newtheorem{definition}{Definition}

\newcommand{\revision}[1]{\textcolor[rgb]{0,0,0}{#1}}

\newcommand{\boldx}{\textbf{x}}

\newcommand{\boldu}{\textbf{u}}

\newcommand{\boldv}{\textbf{v}}

\newcommand{\boldz}{\textbf{z}} 
\newcommand{\boldPhi}{\boldsymbol{\Phi}} 
\newcommand{\K}{\mathcal{K}}
\newcommand{\boldrho}{\boldsymbol{\rho}}

\newcommand{\Ball}{\mathbb{B}}

\newcommand{\Safe}{\mathcal{S}}
\newcommand{\BR}{\text{BR}}

\IEEEoverridecommandlockouts                              % This command is only needed if 
\title{\LARGE \bf 
Robustly Constrained Dynamic Games \\
for Uncertain Nonlinear Dynamics
}
\usepackage{ifthen}
\newboolean{anonymize}

\setboolean{anonymize}{false}
\ifthenelse{\boolean{anonymize}}{
  \author{
  }%
}{
\author{Shuyu Zhan$^{1\star}$ \and Chih-Yuan Chiu$^{2\star}$ \and Antoine P. Leeman$^{3}$\and Glen Chou$^{4}$
\thanks{$^{\star}$Equal contribution.}
\thanks{$^{124}$Georgia Institute of Technology, Schools of $^{1}$Interactive Computing ({\tt\small szhan45@gatech.edu}), $^{2}$Electrical and Computer Engineering ({\tt\small cyc@gatech.edu}), $^{4}$Aerospace Engineering and Cybersecurity \& Privacy ({\tt\small chou@gatech.edu})}
\thanks{$^{3}$Institute for Dynamic Systems and Control, ETH Zürich ({\tt\small aleeman@ethz.ch}).}
\thanks{This work is supported by the European Space Agency OSIP 4000133352.}
}}
\begin{document}

\maketitle
\thispagestyle{empty}
\pagestyle{empty}

%%%%%%%%%%%%%%%%%%%%%%%%%%%%%%%%%%%%%%%%%%%%%%%%%%%%%%%%%%%%%%%%%%%%%%%%%%%%%%%%

% \input{0_Outline}
\begin{abstract}
We propose a novel framework for robust dynamic games with nonlinear dynamics corrupted by state-dependent additive noise, and nonlinear agent-specific and shared constraints. Leveraging system-level synthesis (SLS), each agent designs a nominal trajectory and a causal affine error feedback law to minimize their own cost while ensuring that its own constraints and the shared constraints are satisfied, even under worst-case noise realizations. Building on these nonlinear safety certificates, we define the novel notion of a robustly constrained Nash equilibrium (RCNE). We then present an Iterative Best Response (IBR)-based algorithm that iteratively refines the optimal trajectory and controller for each agent until approximate convergence to the RCNE. We evaluated our method on simulations and hardware experiments involving large numbers of robots with high-dimensional nonlinear dynamics, as well as state-dependent dynamics noise. Across all experiment settings, our method generates trajectory rollouts which robustly avoid collisions, while a baseline game-theoretic algorithm for producing open-loop motion plans failed to generate trajectories that satisfy constraints.
\end{abstract}

\section{Introduction}
\label{sec: Introduction}
% % 

To operate reliably and efficiently, autonomous robots deployed in real-world applications must design and enact robust motion plans that account for both the future motion of surrounding agents and the impact of unpredictable disturbances on system safety.
Recently, dynamic game theory has emerged as a versatile framework for jointly modeling prediction and motion planning in multi-agent interactions \cite{Laine2023ComputationOfApproximateGeneralizedFeedbackNashEquilibria, Kavuncu2021PotentialiLQR, Bhatt2023EfficientConstrainedMultiAgentTrajectoryOptimizationUsingDynamicPotentialGames, Zhou2016CooperativePursuitwithVoronoiPartitions, Peters2024ContingencyGamesforMultiAgentInteraction}.
The core insight of the dynamic games literature is that the interaction outcome between self-interested agents can be captured 
% through the concept of 
by computing
the \textit{Nash equilibrium}, a strategy profile at which each agent selfishly best-responds to the actions of all other agents while satisfying a prescribed set of constraints.
Across a wide range of robotics applications,
% prior work illustrates that, across a broad range of robotics applications, 
the Nash equilibria of 
% \textit{constrained} 
dynamic games has indeed been shown to describe nuanced
% , constraint-satisfying 
% modes of 
interactions among self-interested agents \textit{with deterministic, noise-free dynamics} \cite{basar1998DynamicNoncooperativeGameTheory, isaacs1954differential, fridovich2020efficient, Bhatt2023EfficientConstrainedMultiAgentTrajectoryOptimizationUsingDynamicPotentialGames, Kavuncu2021PotentialiLQR}.

During real-world deployment, however, robots inevitably encounter sources of uncertainty in their dynamics, such as wind disturbances or friction,
% or dynamics model error, 
which may compromise their safety and constraint satisfaction.
Unfortunately, most existing dynamic game formulations either do not explicitly 
account for
% model the impact of
noisy dynamics \cite{lecleach2020algames, fridovich2020efficient, Bhatt2023EfficientConstrainedMultiAgentTrajectoryOptimizationUsingDynamicPotentialGames, Kavuncu2021PotentialiLQR}, and generate trajectories that fail to robustly satisfy safety constraints 
% during deployment 
(see Sec. \ref{sec: Experiments} of our work),
or 
% assume overly conservative noise models 
prescribe overly conservative or defensive motion plans
which ensure constraint satisfaction while greatly sacrificing computational efficiency \cite{Mitchell2005TimeDependentHJBFormulationOfReachableSets}.
% \frank{Will add citation about computational efficiency.}
% \glen{can we get citations here for both cases (not robust; too conservative)?}

To ensure robust constraint satisfaction in interactive motion planning despite unknown noise realizations, we construct a novel algorithm for multi-agent prediction and trajectory generation which integrates key techniques from the robust control literature within the dynamic games framework. 
By coupling motion planning and robust control, our method designs trajectories and error feedback policies
that achieve robust constraint satisfaction, yet are not overly conservative.
Concretely, 
% our work makes 
we make the following contributions:
\begin{figure}
    \centering
    \includegraphics[width=0.96\linewidth]{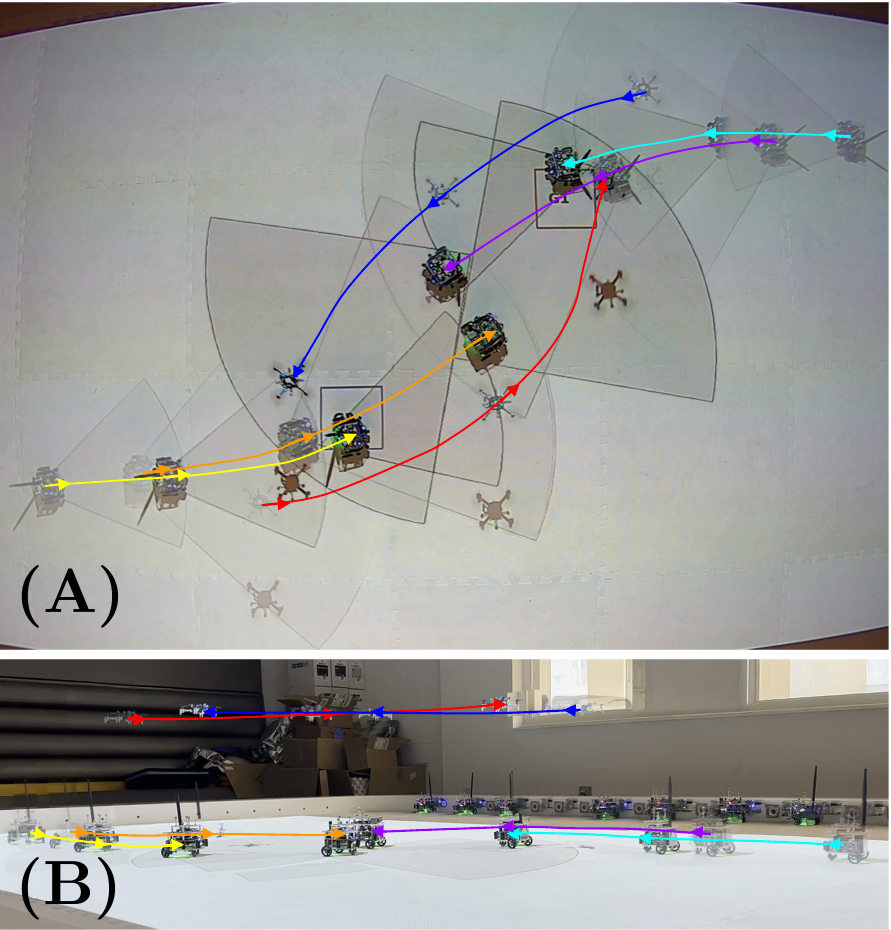}
\caption{
(A) Top and (B) Side views of a hardware experiment in which two heterogeneous robot teams, with 1 quadcopter and 2 ground robots each, navigate within a shared environment. Within each team, one ground robot follows a second, which in turn follows the quadcopter. 
Our method generated interactive motion plans for all robots which
% All robots
obeyed prescribed proximity, line-of-sight, and collision constraints despite 
dynamics noise
% state-dependent noise corruption on their dynamics. 
% For details, 
(see Sec. \ref{subsec: Heterogeneous Team Hardware Experiment}).
% % 
% ~\\
% \frank{To edit}
% ~\\
% % 
% (A)
% Hardware heterogeneous team experiment. 
% Two teams where each team consists of one quadcopter and two ground robots. 
% The quadcopter is modeled as a single integrator with inflated radius $0.1\,\text{m}$, and the ground robots are modeled as 3D Dubins cars with radii 0.06m. One ground robot follows quadcopter and another follows the quadcopter's follower. 
% Followers of quadcopters are subject to proximity and line-of-sight constraints of $0.75\,\text{m}$ and $90^{\circ}$, while followers of unicycles must satisfy $0.5\,\text{m}$ proximity and $60^{\circ}$ field-of-view constraints. 
% The figure shows top and side views of trajectories with initial, middle and terminal conditions of the team, 
% demonstrating that the synthesized controller keeps all agents within their robust error bounds and enforces safety constraints in real-world execution.
}
\vspace{-7mm}
\label{fig:heter_team_hardware}
\end{figure}
\begin{itemize}
    % \vspace{-3mm}
    \item \textbf{Robust Dynamic Games via System Level Synthesis (SLS)}: We extend the \textit{system-level synthesis} (SLS) framework for robust control to a class of multi-agent dynamic games with nonlinear dynamics corrupted with state-dependent additive noise, agent-specific constraints, and shared team constraints. Leveraging recent robust nonlinear SLS results, we jointly optimize the nominal trajectories, 
    % the SLS-parameterized 
    robust controllers, and a 
    % conservative 
    linearization error upper bound across all agents. We obtain tractable safety certificates
    % upper-bound constraint values along error tubes directly in $(z,v,\boldPhi)$ for both $g_t^i$ and $h_t$, 
    through A
    % linear-SLS
    SLS-style robust constraint satisfaction guarantees.
    
    \item 
    \textbf{Formulation and Computation of the Robustly Constrained Nash Equilibrium (RCNE)}: We introduce the novel concept of a \textit{robustly constrained Nash equilibrium} (RCNE), which describes interactions 
    % among strategic agents 
    in which each agent
    % designs a nominal state-control trajectory, robust controller, and linearization error upper bound which
    best responds to all other agents' actions while robustly satisfying a set of prescribed constraints.
    % by combining the SLS parameterization with the above certificates. 
    We develop a novel Iterative Best Response (IBR)-based algorithm to compute the RCNE, which greatly exceeds centralized approaches (e.g., \cite[Alg. 1]{Kavuncu2021PotentialiLQR}) in computational efficiency when applied to the multi-agent robust games studied in this work (Remark \ref{Remark: Compare Fast SLS on nominal problem}). 
    % On $N$-agent, $T$-horizon dynamic games with average agent state and control dimensions given by $\bar n$ and $\bar m$, respectively, our approach yields a computational 
    % % . Each best response step alternates a Riccati-style feedback update with a trajectory update.
    % % yielding 
    % % per-iteration 
    % complexity of $\mathcal{O}(T^2 N (\bar n + \bar m)^3)$ 
    % , while a centralized implementation as in \cite[Alg. 1]{Kavuncu2021PotentialiLQR} would yield a per-iteration complexity of $\mathcal{O}(T^2 N^3 (\bar n + \bar m)^3)$ (Remark \ref{Remark: Compare Fast SLS on nominal problem}).
    
    \item 
    \textbf{Empirical validation}: On a broad range of multi-robot tasks, our approach produced trajectories and controllers that enabled multi-robot teams to interact while avoiding collisions and high-disturbance regions.
    % , and reduce terminal error.
    % while improving safety relative to a non-robust baseline \cite{lecleach2020algames}. 
    Our method scales to 
    % multi-agent interactions 
    interactive scenarios
    involving up to 24 robots, and 
    can accommodate heterogeneous robot teams with multiple air and ground vehicles.
    Our method generated motion plans which satisfied all prescribed constraints across all simulation outcomes, while a dynamic games baseline produced open-loop trajectories which often violated constraints
    % The motion plans constructed by our method satisfy constraints across all simulation outcomes, while the trajectories generated a non-robust, dynamic games baseline often failed to maintain safety 
    % in the presence of
    under
    additive dynamics noise.
\end{itemize}

% \paragraph{Notation} 
\noindent
\textbf{Notation}:
Given integers $N_1, N_2$ with $0 \leq N_1 < N_2$, we define $[N_1, N_2] := \{N_1, \cdots, N_2\}$ and $[N_2] := \{1, \cdots, N_2\}$.
Given $S \subseteq \R^n$ and $M \in \R^{m \times n}$, we define $MS := \{Mx: x \in S \}$.
Given $M_1, \cdots, M_K \in \R^{m \times n}$, we denote by $\diag\{M_1, \cdots, M_K\} \in \R^{mK \times nK}$ the block diagonal matrix with $M_k$ as the $k$-th 
% $m \times n$ 
block, 
% for each 
$\forall k \in [K]$. We denote the horizontal 
% (resp., vertical) 
concatenation of $M_1, \cdots, M_K$ by $\begin{bmatrix}
    M_1 & \cdots & M_K
\end{bmatrix} \in \R^{m \times Kn}$.
% (resp., $[M_1; \cdots; M_K] \in \R^{mK \times n}$). 
For any $m, n \in \N$, we denote the $m \times m$ identity matrix by $I_m$ and the $m \times n$ zero matrix by $O_{m \times n}$, with subscripts omitted when dimensions are clear from context. 
% Given $A \in \R^{m \times n}$, we denote by $A^\dagger \in \R^{n \times m}$ the Moore-Penrose pseudoinverse of $A$.
Given $m \in \N \cup \{+\infty\}$ and $k \in \N$, we denote by $\Ball_m^k$ the unit $m$-norm ball in $\R^k$. 
$\nabla^2$ and $\nabla$ respectively refer to the Hessian matrix and (vertical) gradient vector of a function, taken with respect to all variables.

\section{Related Works}
\label{sec: Related Works}

% \frank{
% Related work:
% \begin{enumerate}
%     \item Dynamic games literature, cite broadly. INCLUDE ILQGAMES and ALGames.
%     % 
%     \item Dynamic games literature, cite open-loop potential games specifically. Mention that they don't discuss robustness w.r.t. noise, while we discuss robustness w.r.t. state-dependent noise.
%     % 
%     \item There are games that consider stochasticity consider uncertainty but often not robustness (cite belief space paper, papers on stochastic games, and GTP-SLAM paper). 
%     % 
%     \item Robust control literature and its connection to zero-sum dynamic games: (1) Only 2 players, typically; (2) too conservative.
%     % 
%     \item (Technical level) SLS---Not multi-agent; ours is the first multi-agent SLS work to the best of the authors' knowledge.
%     % 
%     \item (Algorithmically) Many works use IBR, but ours guarantees robust constraint satisfaction (cite lexigraphic games paper and autonomous racing papers).
% \end{enumerate}
% }

Our work draws from and contributes to the rich literature on game-theoretic multi-agent motion prediction and planning. In particular, \cite{fridovich2020efficient, Laine2023ComputationOfApproximateGeneralizedFeedbackNashEquilibria} designed state feedback policies under the feedback information setting, while \cite{lecleach2020algames, Kavuncu2021PotentialiLQR, Bhatt2023EfficientConstrainedMultiAgentTrajectoryOptimizationUsingDynamicPotentialGames} computed open-loop, interactive multi-agent trajectories. However, whereas \cite{fridovich2020efficient, Laine2023ComputationOfApproximateGeneralizedFeedbackNashEquilibria, lecleach2020algames, Kavuncu2021PotentialiLQR, Bhatt2023EfficientConstrainedMultiAgentTrajectoryOptimizationUsingDynamicPotentialGames} all describe agent motion via deterministic dynamics models, our work considers agent dynamics corrupted by state-dependent noise. Moreover, for each agent, we design a trajectory and controller that satisfy safety constraints despite worst-case noise realizations.

Many works in the game-theoretic motion planning literature incorporate dynamics or agent intent uncertainty within their models. For instance, \cite{Schwarting2021StochasticDynamicGamesinBeliefSpace, Peters2024ContingencyGamesforMultiAgentInteraction} describe interactions with agents who have uncertain intent, while \cite{Zhou2016CooperativePursuitwithVoronoiPartitions} 
% Willman1969FormalSolutionsStochasticPursuitEvasionGames} 
presents dynamic zero-sum games-based methods to perform motion prediction and planning in the presence of an adversarial agent. 
% Meanwhile, 
\cite{Spica2020GameAutonomousRacing, Chiu2022GTPSLAM} present game-theoretic frameworks, with noise-corrupted dynamics, in which agents jointly perform multi-agent prediction, planning, and state estimation. 
However, unlike 
% \cite{Schwarting2021StochasticDynamicGamesinBeliefSpace, Peters2024ContingencyGamesforMultiAgentInteraction}, 
\cite{Schwarting2021StochasticDynamicGamesinBeliefSpace, Peters2024ContingencyGamesforMultiAgentInteraction, 
% Willman1969FormalSolutionsStochasticPursuitEvasionGames, 
Zhou2016CooperativePursuitwithVoronoiPartitions},
which capture \textit{uncertain or adversarial agent intent}, 
we introduce a robust prediction and motion planning framework which guarantees safety despite \textit{uncertainty in agent dynamics}. 
% Moreover, the zero-sum paradigm used in \cite{Willman1969FormalSolutionsStochasticPursuitEvasionGames, Zhou2016CooperativePursuitwithVoronoiPartitions} is typically restricted to adversarial interactions, while our approach accommodates scenarios in which neighboring agents operate under noise-corrupted dynamics, but do not necessarily have \textit{adversarial intent}. 
\cite{Spica2020GameAutonomousRacing, Chiu2022GTPSLAM, Neto2025SLSBRD} consider noise-corrupted agent dynamics but design open-loop trajectories without guarantees of robust constraint satisfaction. In contrast, our method designs trajectories and controllers for noise-corrupted nonlinear systems which provably satisfy prescribed constraints regardless of the noise realization. 
On a technical level, our method  parameterizes robust controllers via system-level synthesis (SLS) \cite{Anderson2019SLS}. Under linear dynamics, multi-agent control via SLS has been studied in MPC formulations \cite{Alonso2023Part1DistributedLocalizedMPC,
Chen2022RobustMPCTimeDelaySystems},
and games \cite{Neto2025SLSBRD}. Meanwhile, \cite{Leeman2025RobustNonlinearOptimalControlviaSLS, Leeman2023CDCRobustOptimalControlforNonlinearSystemswithParametricUncertaintiesviaSLS} extend SLS to nonlinear dynamics but only under affine constraints. Our framework extends 
% \cite{Leeman2025RobustNonlinearOptimalControlviaSLS}
\cite{Neto2025SLSBRD, Alonso2023Part1DistributedLocalizedMPC, 
Chen2022RobustMPCTimeDelaySystems, 
Leeman2025RobustNonlinearOptimalControlviaSLS, Leeman2023CDCRobustOptimalControlforNonlinearSystemswithParametricUncertaintiesviaSLS}
to nonlinearly constrained multi-agent
dynamic games.
% multi-agent game-theoretic setting.
% \antoine{Since we have added the CDC paper, I have replaced the safety filter paper here. It may look a bit weird to have too many self-citations.}
\section{Preliminaries}
\label{sec: Preliminaries}

% (Based on material from Shuyu's Overleaf draft and Dr. Negar Mehr's and Maulik's papers)

% \subsection{Problem Statement}
% \label{subsec: Problem Statement}

% In this section, we introduce core concepts in dynamic game theory (Sec. \ref{subsec: Dynamic Games Setup}) and system-level synthesis (Sec. \ref{subsec: System Level Synthesis}), which we then utilize to formally pose our problem statement (PS) as agent-level and system-level optimization programs (Sec. \ref{subsec: Robust Synthesis for Multi-Agent Games}).

% In this section, we present our multi-agent system 
% (Sec. \ref{subsec: Multi-Agent System Formulation}),

% \subsection{Multi-Agent System Formulation}
% \label{subsec: Multi-Agent System Formulation}

Consider an $N$-agent, $T$-stage discrete-time dynamic game $\G$, in which $x_t^i \in \R^{n_i}$ and $u_t^i \in \R^{m_i}$ respectively represent the \textit{state} and \textit{control} vector of each agent $i \in [N]$ at each time $t \in [0, T]$. 
We use $x_t := (x_t^1, \cdots, x_t^N) \in \R^n$ and $u_t := (u_t^1, \cdots, u_t^N) \in \R^m$ to respectively denote the \textit{system state} and \textit{system control} at each time $t$, where $n := \sum_{i=1}^N n_i$ and $m := \sum_{i=1}^N m_i$.
\revision{We also write $\boldx^i := (x_0^i, \cdots, x_T^i)$ and $\boldu^i := (u_0^i, \cdots, u_T^i)$ for each agent $i \in [N]$.}
We also write $\boldx := (x_0, \cdots, x_T) \in \R^{n(T+1)}$ and $\boldu := (u_0, \cdots, u_{T-1}) \in \R^{mT}$ to denote the system state trajectory and system control trajectory, respectively. 
Moreover, we set $u_T^i = 0 \ \forall i \in [N]$. 
% , and that $u_T = 0$.

Each agent $i \in [N]$ is associated with an initial state $x_0^i = \bar x_0^i \in \R^{n_i}$, as well as noise-corrupted dynamics given by \footnote{Our experiments also model uncertainty in $x_0$, which can be incorporated into the dynamics noise. For details, see \cite{Leeman2023CDCRobustOptimalControlforNonlinearSystemswithParametricUncertaintiesviaSLS}.}:
\begin{align} \label{Eqn: Dynamics, Agent i}
    x_{t+1}^i &= f_t^i(x_t^i,u_t^i) + E_t^i(x_t^i) w_t^i,
\end{align}
where $f_t^i: \R^{n_i} \times \R^{m_i} \ra \R^{n_i}$ denotes the dynamics map,
$w_t^i \in \R^{n_i}$ denotes additive noise,
% $w_t^i \in \{w \in \R^{n_i}: \Vert w \Vert_\infty \leq 1 \}$ denotes unit $\infty$-norm disturbance in $\R^{n_i}$, 
and $E_t^i(x_t^i) \in \R^{n_i} \times \R^{n_i}$ describes state-dependent disturbance scaling. 

\begin{remark} \label{Remark: Extension to Parametric Uncertainty}
Our dynamics model \eqref{Eqn: Dynamics, Agent i} can be straightforwardly adapted to incorporate parametric uncertainty in the agent dynamics, as considered in \cite{Leeman2023CDCRobustOptimalControlforNonlinearSystemswithParametricUncertaintiesviaSLS}.
\end{remark}

As specified below, each agent $i \in [N]$ aims to minimize their own overall cost, 
% $J^i(\boldx, \boldu)$, where $J^i: \R^{(n+m)T} \ra \R$, 
while satisfying both their individual constraints and a set of team constraints shared among the $N$ agents. 
Each agent has stage-wise costs $L_t^i: \R^n \times \R^m \ra \R$ and an overall cost $J^i: \R^{n(T+1)} \times \R^{mT} \ra \R$, given by:
\begin{align} \label{Eqn: Overall Cost J i, as sum of Stage-wise costs L ti}
    J^i(\boldx, \boldu) &:= \textstyle\sum_{t=0}^T L_t^i(x_t, u_t).
\end{align}
% The \textit{overall cost function} for each agent $i \in [N]$, denoted by $J^i: \R^{n(T+1)} \times \R^{mT} \ra \R$ and defined as follows, is defined by summing Agent $i's$ 

We note that for each agent $i \in [N]$ and time $t \in [T]$, $J^i$ and $L_t^i$ depend in general on \textit{all agents'} control decisions and resulting trajectories.
% Concretely, the overall cost $J^i$ of each agent is given by the sum of stage-wise cost terms $L_t^i: \R^{n+m} \ra \R, \ \forall t \in [T]$, i.e.,: 
% and $L_T^i: \R^n \ra \R$, i.e.,:
% \begin{align}
%     J^i(\boldx, \boldu) &= \sum_{t=0}^{T-1} L_t^i(x_t^i, u_t^i, x_t^{-i}, u_t^{-i}).
%     % + L_T^i(x_T^i, x_T^{-i})
% \end{align}
Constraints specific to each agent $i$ at each time $t$ are defined by a collection of $n_g^i$ scalar constraint functions $g_{t,k}^i: \R^{n_i} \times \R^{m_i} \ra \R$ and $g_{T,k}^i: \R^{n_i} \times \R^{m_i} \ra \R$ across all $k \in [n_g^i]$, as shown below:
\begin{align} \label{Eqn: Individual Constraints}
    g_{t,k}^i(x_t^i, u_t^i) &\leq 0, \hspace{5mm} \forall \ t \in [T-1], k \in [n_g^i].
    % \\ \label{Eqn: Individual Constraints, at time t}
    % g_{T,k}^i(x_T^i) &\leq 0, \hspace{5mm} \forall \ k \in [n_g^i].
\end{align}
Moreover, constraints shared by the $N$ agents at each time $t \in [T]$ are defined by a collection of $n_h$ scalar constraint functions $h_{t,k}: \R^n \times \R^m \ra \R$ and $h_{T,k}: \R^n \ra \R$ across all $k \in [n_h]$,
as shown below:
\begin{align} \label{Eqn: Shared Team Constraints}
    h_{t,k}(x_t, u_t) &\leq 0, \hspace{5mm} \forall \ t \in [T-1], k \in [n_h].
    % \\ \label{Eqn: Shared Team Constraints, at time t}
    % h_{T,k}(x_T) &\leq 0, \hspace{5mm} \forall \ k \in [n_h].
\end{align}
% For the case $t = T$ in \eqref{Eqn: Individual Constraints}-\eqref{Eqn: Shared Team Constraints}, we adopt the convention that $u_T^i = 0$ for each agent $i \in [N]$.

In our work, we 
% restrict our analysis to the class of constrained \textit{dynamic potential games}
% We 
assume the dynamic game $\G$ under study is a \textit{constrained dynamic potential game} with a corresponding potential function $P: \R^{n(T+1)} \times \R^{mT} \ra \R$, as described in \cite{Bhatt2023EfficientConstrainedMultiAgentTrajectoryOptimizationUsingDynamicPotentialGames, Zanardi2021LexicographicGames} and defined below.
Dynamic potential games provide a versatile and widely used modeling framework for motion prediction and trajectory generation in multi-agent robotics and controls applications \cite{Bhatt2023EfficientConstrainedMultiAgentTrajectoryOptimizationUsingDynamicPotentialGames, Kavuncu2021PotentialiLQR, Spica2020GameAutonomousRacing}.

\begin{definition} \label{Def: Dynamic Potential Game}
The game $\G$ is called a \textit{constrained dynamic potential game} if there exists a \textit{potential function} $P: \R^{n(T+1)} \times \R^{mT} \ra \R$ such that, for any feasible 
% nominal 
trajectories $(\boldx, \boldu), (\hat \boldx, \hat \boldu) \in \R^{n(T+1)} \times \R^{mT}$, and any agent $i \in [N]$:
\begin{align}
    &P(\hat \boldx^i, \boldx^{-i}, \hat \boldu^i, \boldu^{-i}) - P(\boldx, \boldu) \\ \nonumber
    = \ &J^i(\hat \boldx^i, \boldx^{-i}, \hat \boldu^i, \boldu^{-i}) - J^i(\boldx, \boldu).
\end{align}
In words, $P$ tracks the change in each agent $i$'s cost $J^i$ corresponding to that agent's unilateral deviation from one state-control trajectory to another.
\end{definition}

\begin{assumption} \label{Assump: Dynamic Potential Game}
$\G$ is a constrained dynamic potential game.
\end{assumption}

% \paragraph{Key technical assumptions} 

% \subsection{System Level Synthesis for Multi-Agent Systems}
% \label{subsec: System Level Synthesis for Multi-Agent Systems}

\section{Robustly Constrained Nash Equilibrium (RCNE) and Problem Formulation}
\label{sec: Robustly Constrained Nash Equilibrium (RCNE) and Problem Formulation}

Here, we introduce core concepts in \textit{system-level synthesis} to design robust controllers for dynamic potential games (Sec. \ref{subsec: Robust Controller Synthesis for Dynamic Games via SLS}), and formulate a novel notion of Nash equilibrium that accounts for robust constraint satisfaction in addition to optimal agent actions (Sec. \ref{subsec: Robustly Constrained Nash Equilibrium (RCNE)}). We 
% then 
formally present our problem statement at the end of Sec. \ref{subsec: Robustly Constrained Nash Equilibrium (RCNE)}.
% Then, in Sec. \ref{subsec: Robustly Constrained Nash Equilibrium (RCNE)}, we formally pose our problem statement from Sec. \ref{sec: Introduction} as agent-level and system-level optimization programs.

\subsection{Robust Controller Synthesis for Dynamic Games via SLS}
\label{subsec: Robust Controller Synthesis for Dynamic Games via SLS}

To enable robust controller synthesis for dynamic games with noise-corrupted nonlinear dynamics, we leverage the 
system level synthesis (SLS)-based
% robust nonlinear optimal control 
methods in \cite{Leeman2025RobustNonlinearOptimalControlviaSLS}.
Within the SLS framework, each agent $i \in [N]$ synthesizes a safe controller for the noise-corrupted dynamics \eqref{Eqn: Dynamics, Agent i} based on the decomposition of
%for decomposing 
a nonlinear system into the sum of a \textit{nominal} nonlinear system and a set of linear time-varying (LTV) error dynamics.
% , which we present below in the context of a multi-agent system. 
% In particular, given a noise-corrupted nonlinear system with bounded disturbances, \cite{Leeman2025RobustNonlinearOptimalControlviaSLS} utilizes the \textit{system-level synthesis} (SLS) framework introduced in \cite{Anderson2019SLS} to jointly synthesize a nominal trajectory, error feedback policy, and linearization bounds.
Concretely, each agent $i \in [N]$ designs (a) a nominal state trajectory $\boldz^i := (z_0^i, \cdots, z_T^i) \in \R^{n_i(T+1)}$, nominal control trajectory $\boldv^i := (v_0^i, \cdots, v_{T-1}^i) \in \R^{m_i T}$,
satisfying the \textit{nominal dynamics}:
\begin{align} \label{Eqn: Nominal, Dynamics, Agent i}
    z_{t+1}^i &= f_t^i(z_t^i, v_t^i), \hspace{5mm} \forall t \in [0, T-1],
\end{align}
and (b) a causal affine error feedback law of the form:
\begin{align} \label{Eqn: Causal affine error feedback law, original, u in terms of v, K, and state error}
    u_t^i = \pi_t^i(x_{0:t}^i) &= v_t^i + \textstyle\sum_{\tau=0}^{t-1} K_{t-1,\tau}^i (x_{t-\tau}^i - z_{t-\tau}^i).
\end{align}
% Above, \eqref{Eqn: Causal affine error feedback law, original, u in terms of v, K, and state error} can be compactly written as $\boldu^i = \boldv^i + \K^i (\boldx^i - \boldz^i)$,
% % \begin{align}
% %     \boldu^i &= \boldv^i + \K^i (\boldx^i - \boldz^i),
% % \end{align}
% where $\K^i \in \R^{m_i T \times n_i T}$, the \textit{state feedback matrix} of agent $i$, is defined as the matrix whose $(t,\tau)$-th $m_i \times n_i$ block, for each $t,\tau \in [T]$ equals $K_{t-1,\tau-1}^i$ if $t \geq \tau$, and equals the zero matrix otherwise. 

Consider the linearization of \eqref{Eqn: Dynamics, Agent i} about Agent $i$'s nominal state-control trajectory $(\boldz^i, \boldv^i)$. 
% $\boldz := (z_1, \cdots, z_T) \in \R^{n(T+1)}$ and nominal control trajectory $\boldv := (v_1, \cdots, v_T) \in R^{mT}$, which satisfy the nominal dynamics $z_{t+1}^i = f_t^i(z_t^i, v_t^i)$ for each agent $i \in [N]$. 
We define the trajectory and control error, $\forall i \in [N]$, $t \in [T]$, by $\Delta x_t^i := x_t^i - z_t^i \in \R^{n_i}$ and $\Delta u_t^i := u_t^i - v_t^i \in \R^{m_i}$ respectively. Next, we define the Jacobian matrices $A_t^i: \R^{n_i}\times \R^{m_i} \ra \R^{n_i \times n_i}$ and $B_t^i: \R^{n_i}\times \R^{m_i} \ra \R^{n_i \times m_i}$ for the dynamics \eqref{Eqn: Dynamics, Agent i} 
% and write $e_t := (e_t^1, \cdots, e_t^N) \in \R^n$ for each time $t \in [T]$, $e^i := (e_1^i, \cdots, e_T^i) \in \R^{n_i T}$ for each agent $i \in [N]$, and $e := (e_1, \cdots, e_T) \in \R^{n(T+1)}$. 
and the
% remainders $r_t^i: \R^{2(n_i + m_i)} \ra \R^{n_i}$ 
disturbance $d_t^i$, as follows. 
We note that $d_t^i$ describes the impact of both the linearization error and the dynamics noise:
% and $r_{E,t}^i: \R^{2n_i} \ra \R^{n_i \times n_i}$ 
\begin{align}
    \nonumber
    A_t^i(z_t^i, v_t^i) &:= \frac{\partial f_t^i}{\partial x_t^i}(z_t^i, v_t^i), \\ \nonumber
    B_t(z_t^i, v_t^i) &:= \frac{\partial f_t^i}{\partial u_t^i}(z_t^i, v_t^i), \\ \label{Eqn: d ti, def}
    % r_t^i(x_t^i, u_t^i, z_t^i, v_t^i)
    d_t^i &:= f_t^i(x_t^i, u_t^i) - f_t^i(z_t^i, v_t^i) -
    % \frac{\partial f_t^i}{\partial x_t^i}
    A_t^i(z_t^i, v_t^i) \Delta x_t^i \\ \nonumber
    &\hspace{5mm} - 
    % \frac{\partial f_t^i}{\partial u_t^i}
    B_t^i
    (z_t^i, v_t^i) \Delta u_t^i + E_t^i(x_t^i) w_t^i.
    % , \\
    % r_{E,t}^i(x_t^i, z_t^i) &:= E_t^i(x_t^i) - E_t^i(z_t^i).
\end{align}
We can then express \eqref{Eqn: Dynamics, Agent i} in terms of the \textit{nominal dynamics} \eqref{Eqn: Nominal, Dynamics, Agent i} and the following LTV error dynamics:
\begin{subequations} \label{Eqn: Error dynamics and State error feedback, Delta x ti and Delta u ti}
\begin{align} 
    % z_{t+1}^i &:= f(z_t^i, v_t^i), \\
    \label{Eqn: Error dynamics, Delta x ti}
    \Delta x_{t+1}^i &= A_t(z_t^i, v_t^i) \Delta x_t^i + B_t(z_t^i, v_t^i) \Delta u_t^i + d_t^i, \\ \label{Eqn: State error feedback, Delta u ti}
    \Delta u_t^i &= \textstyle\sum_{\tau=0}^{t-1} K_{t-1,\tau}^i \Delta x_{t-\tau}^i,
\end{align}
\end{subequations}
where \eqref{Eqn: State error feedback, Delta u ti} rewrites \eqref{Eqn: Causal affine error feedback law, original, u in terms of v, K, and state error} using notation of the form $\Delta x_t^i$ and $\Delta u_t^i$. 
We stack the state and control error as $e_t^i := (\Delta x_t^i, \Delta u_t^i) \in \R^{n_i+m_i}$.

A core tenet of the SLS framework is that, for each agent $i \in [N]$, all causal affine error feedback gains 
% (as defined in \eqref{Eqn: Causal affine error feedback law, original, u in terms of v, K, and state error}):
% $\K := \{K_{t,\tau}^i: t,\tau \in [0,T-1], t \geq \tau \}$ 
\begin{align} \label{Eqn: K i}
    \K^i := \{K_{t,\tau}^i: t,\tau \in [0,T-1], t \geq \tau \}
\end{align}
can be parameterized by a set of \textit{system response matrices}:
\begin{align} \label{Eqn: Phi i}
    \boldPhi^i := &\{\boldPhi_{t,\tau}^i \revision{= [(\boldPhi^i)_{t,\tau}^x; (\boldPhi^i)_{t,\tau}^u]} \in \R^{(n_i+m_i) \times n_i}: \\ \nonumber
    &\hspace{1cm} i \in [N], t, \tau \in [0, T-1], t \geq \tau \}
\end{align}
satisfying \eqref{Eqn: Phi, Equality Constraints for feasibility and causality, for each t, tau, x} (see \cite[Sec. 3]{Leeman2025RobustNonlinearOptimalControlviaSLS} 
% App. \ref{subsec: App, Proof of Proposition on Robust Constraint Satisfaction} and 
\cite[Sec. 2]{Anderson2019SLS}). 
We formally present this parameterization below in Prop. \ref{Prop: SLS, Using Phi  to Parameterize K}. 
% For brevity, we omit the proof of Prop. 1, which follows from the error dynamics analysis presented in \cite[Sec. 3B]{Leeman2025RobustNonlinearOptimalControlviaSLS}.

% which follows from the error dynamics analysis provided in \cite[Sec. 3B]{Leeman2025RobustNonlinearOptimalControlviaSLS}, formalizes the parametrization of the affine controller \eqref{Eqn: State error feedback, Delta u ti} for the dynamics \eqref{Eqn: Error dynamics, Delta x ti} in terms of a \textit{system response} $\Phi^i$ characterized below.

\begin{proposition}(\textbf{Adapted from \cite[Sec. 3B]{Leeman2025RobustNonlinearOptimalControlviaSLS}}) \label{Prop: SLS, Using Phi to Parameterize K}
Suppose each agent $i \in [N]$ 
% and horizon $T$,  
is associated with dynamics \eqref{Eqn: Dynamics, Agent i}, a nominal trajectory $(\boldz^i, \boldv^i)$, and linearized error dynamics \eqref{Eqn: Error dynamics, Delta x ti}. 
Then for each agent $i \in [N]$, the following hold true:
\begin{enumerate}
    \item Let a causal affine error feedback law \eqref{Eqn: State error feedback, Delta u ti} 
    %characterized by 
    %$\K^i$, of the form \eqref{Eqn: K i}, 
    % $\K^i := \{K_{t,\tau}^i: t,\tau \in [0, T-1], t \geq \tau\}$ 
    be given. Then there exists a system response
    $\boldPhi^i$
    % $\boldPhi^i := \{\Phi_{t, \tau}^i \in \R^{(n_i+m_i) \times n_i}: t, \tau \in [0, T-1], t \geq \tau \}$ 
    % for agent $i \in [N]$
    of the form \eqref{Eqn: Phi i}, such that $\forall t \in [0,T-1]$, the rollout error $e_t^i = (\Delta x_t^i, \Delta u_t^i)$ satisfying \eqref{Eqn: Error dynamics and State error feedback, Delta x ti and Delta u ti} is a linear function of the disturbance terms $\{d_t^i: i \in [N], t \in [T] \}$, given by:
    \begin{align} \label{Eqn: Disturbance Propagation, e from Phi and d}
        e_t^i 
        % = \begin{bmatrix}
        %     \Delta x_t^i \\ \Delta u_t^i
        % \end{bmatrix}
        = 
        \begin{bmatrix}
            \sum_{\tau=0}^{t-1}  (\boldPhi^i)_{t-1,\tau}^x d_\tau^i \\ \sum_{\tau=0}^{t-1}  (\boldPhi^i)_{t-1,\tau}^u d_\tau^i
        \end{bmatrix} = \sum_{\tau=0}^{t-1} \boldPhi_{t-1,\tau}^i d_\tau^i.
    \end{align}
    % where $e_t^i$ satisfies the LTV error \eqref{Eqn: Error dynamics and State error feedback, Delta x ti and Delta u ti}.

    \item Conversely, let a system response $\boldPhi^i$ of the form \eqref{Eqn: Phi i}, encoding disturbance propagation as in \eqref{Eqn: Disturbance Propagation, e from Phi and d}, be given. If $\boldPhi^i$ satisfies \eqref{Eqn: Phi, Equality Constraints for feasibility and causality, for each t, tau, x} below, then it induces a unique causal affine error feedback of the form \eqref{Eqn: State error feedback, Delta u ti} corresponding to the disturbance propagation described by $\boldPhi^i$:
    % Conversely, any collection $\boldPhi^i$ of the form \eqref{Eqn: Phi i} obeying \eqref{Eqn: Phi, Equality Constraints for feasibility and causality, for each t, tau, x} below induces a unique causal affine error feedback $\K^i$ of the form \eqref{Eqn: K i} corresponding to the disturbance propagation described by $\Phi^i$:
    \begin{subequations} \label{Eqn: Phi, Equality Constraints for feasibility and causality, for each t, tau, x}
    \begin{align} \nonumber
        (\boldPhi^i)_{t+1, \tau}^x &= A_t^i(z_t^i, v_t^i) (\boldPhi^i)_{t, \tau}^x + B_t^i(z_t^i, v_t^i) (\boldPhi^i)_{t, \tau}^u, \\ 
        &\hspace{1cm} \forall t, \tau \in [0, T-2], t \geq \tau, \\
        (\boldPhi^i)_{t, t}^x &= I_{n_i}, \hspace{1cm} \forall t \in [0, T-1].
    \end{align}
    \end{subequations}

\end{enumerate}

\end{proposition}

We now present Prop. \ref{Prop: Robust Constraint Satisfaction}, a refinement of the characterization of the robust constraint satisfaction in \cite[Prop. III.3]{Leeman2025RobustNonlinearOptimalControlviaSLS} to the setting of non-linear constraints. Prop. \ref{Prop: Robust Constraint Satisfaction} requires the following mild assumptions on the boundedness of the states $x_t$, controls $u_t$, noise terms $w_t$, and state-dependent maps $E_t^i(x_t^i) \in \R^{n_i \times n_i}$ in our dynamic game model, as well as the regularity of the dynamics and constraint maps $f_t^i$, $g_t^i$, and $h_t$, across agent and time indices.

\begin{assumption} \label{Assump: Bounded States, Controls, per agent}
% For each agent $i \in [N]$, there exists a compact set $\Omega^i \subset \R^{n_i + m_i}$ such that, for any $(x^i, u^i) \in \R^{n_i + m_i}$ satisfying \eqref{Eqn: Individual Constraints}-\eqref{Eqn: Shared Team Constraints}, we have $(x^i, u^i) \in \Omega^i$. 
Let $\Omega \subset \R^{n + m}$ be the set of all $(x_t, u_t)$ satisfying \eqref{Eqn: Individual Constraints}-\eqref{Eqn: Shared Team Constraints}. We assume $\Omega \subset \R^{n + m}$ is compact.
    % For convenience, we define:
    % \begin{align}
    %     \Omega_x &:= \{ 
    %     \begin{bmatrix}
    %         I_n & O_{n \times m}
    %     \end{bmatrix} 
    %     (x,u): (x,u) \in \Omega \}, \\
    %     \Omega_u &:= \{
    %     \begin{bmatrix}
    %         O_{m \times n} & I_m
    %     \end{bmatrix} 
    %     (x,u): (x,u) \in \Omega \}.
    % \end{align}
\end{assumption}

\begin{assumption} \label{Assump: Dynamics, constraint maps, Hessian bounds}
The dynamics and constraint maps $f_t^i$, $g_t^i$, and $h_t$ are twice continuously differentiable.
\end{assumption}

For each $i \in [N]$, $t \in [T]$, and $k \in [n_i]$, let $f_{t,k}^i: \R^{n_i} \times \R^{m_i}\ra \R$ denote the $k$-th scalar output of the dynamics map $f_t^i$. 
% Similarly, for each $i \in [N]$, $t \in [T]$, we define $g_{t,k}^i: \R^{n_i + m_i} \ra \R$ to be the $k$-th scalar output of the Agent $i$-specific constraint $g_t^i: \R^{n_i + m_i} \ra \R^{n_g^i}$, across indices $k \in [n_g^i]$ and we define $h_{t,k}: \R^{n + m} \ra \R$ to be the $k$-th scalar output of the shard constraint $h_t: \R^{n_i + m_i} \ra \R^{n_h}$, across indices $k \in [n_h]$.
Given Assumption \ref{Assump: Dynamics, constraint maps, Hessian bounds}, we define, 
$\forall i \in [N]$, $t \in [T]$:
\begin{align} \label{Eqn: f tki Hessian bound, denoted mu tki}
    \mu_{t,k}^i &:= \frac{1}{2} \max_{\substack{(\tilde x_t, \tilde u_t) \in \Omega \\ \xi \in \Ball_\infty^{n_i + m_i} }}  \xi^\top \nabla^2 f_{t,k}^i(\tilde x_t^i, \tilde u_t^i) \xi, \hspace{3mm} \forall k \in [n_i], 
    % < \infty.
    \\ \label{Eqn: g tki Hessian bound, denoted chi tki}
    \chi_{t,k}^i &:= \frac{1}{2} \max_{\substack{(\tilde x_t, \tilde u_t) \in \Omega \\ \xi \in \Ball_\infty^{n_i + m_i}}}  \xi^\top \nabla^2 g_{t,k}^i(\tilde x_t^i, \tilde u_t^i) \xi, \hspace{3mm} \forall k \in [n_g^i], 
    % < \infty, 
    \\ \label{Eqn: h tk Hessian bound, denoted psi tki}
    \psi_{t,k} &:= \frac{1}{2} \max_{\substack{(\tilde x_t, \tilde u_t) \in \Omega \\ \zeta \in \Ball_\infty^{n+m}}}  \zeta^\top \nabla^2 h_{t,k}(\tilde x_t, \tilde u_t) \zeta, \hspace{3mm} \forall k \in [n_h].
    % < \infty, 
\end{align}

% Finally, we impose the following bounds on the noise terms $w_t^i$ and the state-dependent maps $E_t^i(x_t^i) \in \R^{n_i \times n_i}$.

\begin{assumption} \label{Assump: Bounded Noise, Lipschitz noise map}
Each noise term $w_t^i$ lies within the $\infty$-norm ball in $\R^{n_i}$, i.e., $w_t^i \in \Ball_\infty^{n_i}$. Moreover, the state-dependent disturbance multiplier functions $E_t^i: \R^{n_i} \times \R^{n_i}$ are Lipschitz.
\end{assumption}

Given Assumption \ref{Assump: Bounded Noise, Lipschitz noise map}, for each $i \in [N]$, $t \in [T]$, $k \in [n_i]$, let $L_{E,t,k}^i > 0$ denote the Lipschitz constant for the $k$-th row of $n_i \times n_i$ matrix $E_t^i$, denoted $E_{t, k}^i: \R^{n_i} \ra \R^{1 \times n_i}$ below. Concretely, for any $x_t^i, z_t^i \in \R^{n_i}$:
\begin{align}
\label{Eqn: L etk i}
    \Vert E_{t,k}^i(x_t^i) - E_{t,k}^i(z_t^i) \Vert_\infty \leq L_{E,t,k}^i \Vert x_t^i - z_t^i \Vert_\infty.
\end{align}

% ~\\
% \frank{Add a remark describing how the above parameters can be estimated in practice (see similar remarks made by Glen in iSLS paper).}
% ~\\
\begin{remark}
    \looseness-1$\mu_{t,k}^i$, $\chi_{t,k}^i$, $\psi_{t,k}$, and $L_{E,t,k}^i$ can be estimated via extreme value theory \cite{Knuth2023StatistcalSafetyRobustnessGuarantees}
    or interval arithmetic \cite{limon2005robust}.
\end{remark}

% Concretely, Prop. \ref{Prop: Robust Constraint Satisfaction} upper bounds the agent-indexed constraint function values $g_{t,k}^i(x_t^i, u_t^i)$ 
% % across indices $t \in [T]$, $i \in [N]$, $k \in [n_g^i]$, 
% and shared constraints $h_{t,k}(x_t, u_t)$, in terms of their values as evaluated on the nominal trajectory $(\boldz, \boldv)$, the dynamics noise weights $E_t^i(z_t^i)$ on the nominal trajectory $(\boldz^i, \boldv^i)$, the magnitude of the error vector $e^i$, and the system response $\boldPhi^i$.
% % corresponding to the each agent's state feedback law $\K^i$. 

% Some preliminary notation for Prop. \ref{Prop: Robust Constraint Satisfaction} are as follows.

We now present Prop. \ref{Prop: Robust Constraint Satisfaction},
% (proof given in App. \ref{subsec: App, Proof of Proposition on Robust Constraint Satisfaction})
the proof of which is given in our ArXiv paper \cite[App. A]{ArXivPaper}.
% (proof given in App. \ref{subsec: App, Proof of Proposition on Robust Constraint Satisfaction}).

\begin{proposition}[\textbf{Robust Constraint Satisfaction}] \label{Prop: Robust Constraint Satisfaction}
\looseness=-1
Suppose nominal trajectories $\boldz = (\boldz^1, \cdots, \boldz^N)$, $\boldv := (\boldv^1, \cdots, \boldv^N)$ and system responses $\boldPhi = \{\boldPhi^i: i \in [N] \}$ satisfying \eqref{Eqn: Phi, Equality Constraints for feasibility and causality, for each t, tau, x} are given. For each $i \in [N]$, $t,\tau \in [T]$ such that $t \geq \tau$, we define $\Lambda_t^i(\boldz, \boldv) \in \R^{n_i \times 2n_i}$, $\Gamma_{t,\tau}(\boldz, \boldv, \boldPhi) \in \R^{n \times 2n}$, $\mu_t^i \in \R^{n_i \times n_i}$, and $L_{E,t}^i \in \R^{n_i \times n_i}$ as follows:
\begin{align} \label{Eqn: Lambda ti}
    \Lambda_t^i(\boldz, \boldv) &:= \begin{bmatrix}
        E_t^i(z_t^i) & \Vert e_t^i \Vert_\infty^2 \mu_t^i + \Vert e_t^i \Vert_\infty L_{E,t}^i
    \end{bmatrix} \\ 
    % \nonumber
    % &\hspace{1cm}\in \R^{n_i \times 2n_i}, \\ 
    \label{Eqn: Gamma t tau}
    \Gamma_{t,\tau}(\boldz, \boldv, \boldPhi) &:= \diag \big\{ \boldPhi_{t,\tau}^1 \Lambda_{t-\tau}^1(\boldz, \boldv), \cdots, \boldPhi_{t,\tau}^N \Lambda_{t-\tau}^N(\boldz, \boldv) \big\} \\ \label{Eqn: mu ti}
    % \nonumber
    % &\hspace{5mm} \in \R^{n \times 2n}, \\ 
    \mu_t^i &:= \diag\big\{ \mu_{t,1}^i, \cdots, \mu_{t,n_i}^i \big\}, 
    % \in \R^{n_i \times n_i}, 
    \\ \label{Eqn: L Eti} 
    L_{E,t}^i &:= \diag\big\{ L_{E,t,1}^i, \cdots, L_{E,t,n_i}^i \big\}.
    % \in \R^{n_i \times n_i}.
\end{align}
If there exist auxiliary \textit{error upper bound} variables $\boldrho^i := (\rho_0^i, \cdots, \rho_{T-1}^i) \in \R^T$ such that the following hold $\forall i \in [N]$:
\vspace{-2mm}
\begin{subequations} 
\label{Eqn: g tki and h tk upper bound, to use in opt}
\begin{align} 
\nonumber
    &g_{t,k}^i(z_t^i, v_t^i) + \sum_{\tau=0}^{t-1} \big\Vert \nabla g_{t,k}^i(z_t^i, v_t^i)^\top \boldPhi_{t-1,\tau}^i \Lambda_{t-1-\tau}(\boldz) \big\Vert_1 \\ \label{Eqn: g tki upper bound, to use in opt}
    &\hspace{5mm} + \chi_{t,k}^i (\rho_t^i)^2 \leq 0, \hspace{3mm} \forall t \in [T], k \in [n_g^i], \\ 
    \nonumber
    &h_{t,k}(z_t, v_t) + \sum_{\tau=0}^{t-1} \big\Vert \nabla h_{t,k}(z_t, v_t)^\top \Gamma_{t-1,\tau}(\boldz, \boldPhi)
    \big\Vert_1 \\ \label{Eqn: h tk upper bound, to use in opt}
    &\hspace{5mm} + \psi_{t,k} \cdot \sum_{j=1}^N (\rho_t^j)^2 \leq 0, \hspace{3mm} \forall t \in [T], k \in [n_h], \\ \nonumber 
    &\sum_{\tau=0}^{t-1} \Vert \boldPhi_{t-1, \tau}^i \begin{bmatrix}
        E_\tau^i(z_\tau^i) & (\rho_\tau^i)^2 \mu_\tau^i + \rho_\tau^i L_{E,\tau}^i
    \end{bmatrix} \Vert_\infty \leq \rho_t^i, \\ \label{Eqn: rho recursive bound, to use in opt}
    &\hspace{5mm} \forall t \in [T-1], \\ \label{Eqn: rho initial bound at t = 0, to use in opt} 
    &\rho_0^i \geq 0.
\end{align}
\end{subequations}
then, for any realization of the noise terms $w_t^i$, we have $g_{t,k}^i(x_t^i, u_t^i) \leq 0$ $\forall k \in [n_g^i]$ and $h_{t,k}(x_t, u_t) \leq 0$ $\forall k \in [n_h]$, for each $i \in [N]$, $t \in [T]$, across all trajectory rollouts $(\boldx, \boldu) \in \R^{n(T+1) + mT}$.
\end{proposition}
\subsection{Robustly Constrained Nash Equilibrium}
\label{subsec: Robustly Constrained Nash Equilibrium (RCNE)}

We now utilize the SLS-based parameterization of the affine error feedback for the LTV system (Sec. \ref{subsec: Robust Controller Synthesis for Dynamic Games via SLS}) to define the \textit{robustly constraint-satisfying Nash equilibrium} (RCNE) solution to our dynamic game (Def. \ref{Def: Robustly Constrained Nash Equilibrium (RCNE)}). In words, the RCNE which characterize a set of steady-state agent trajectories and controls at which each agent acts optimally with respect to all other agents' actions, while ensuring constraint satisfaction even under worst-case noise realizations.

% ~\\
% \frank{To self: Incorporate $\boldrho$ variables into discussion below.}
% ~\\

% To begin, 
% we collect the nominal state-control trajectories $(\boldz^i, \boldv^i)$ and system responses $\boldPhi^i$ across all agents $i \in [N]$ to form $\boldz := (\boldz^1, \cdots, \boldz^N)$, $\boldv := (\boldv^1, \cdots, \boldv^N)$, and $\boldPhi := (\boldPhi^1, \cdots, \boldPhi^N)$.
% we 
We first 
define the set of \textit{robustly constraint-satisfying} agent nominal trajectories and system responses
% , denoted $\Safe$ below, 
by:
\begin{align} 
    \nonumber
    \Safe := \big\{ (\boldz, \boldv, \boldPhi, \boldrho): &(\boldz^i, \boldv^i, \boldPhi^i, \boldrho^i) \text{ satisfy } \\
    &\hspace{5mm} \eqref{Eqn: Nominal, Dynamics, Agent i}, \eqref{Eqn: Phi, Equality Constraints for feasibility and causality, for each t, tau, x}, \eqref{Eqn: g tki and h tk upper bound, to use in opt}, \forall \ i \in [N] \big\},
\end{align}
where $\boldz$, $\boldv$, and $\boldPhi$ are as defined in Sec. \ref{subsec: Robust Controller Synthesis for Dynamic Games via SLS}, and $\boldrho := \{\boldrho^i: i \in [N] \}$. Next, similar to \cite{Leeman2024FastSLS}, to control the uncertainty propagation, we introduce the following augmented cost $\tilde J^i$ for each agent $i \in [N]$, which appends a convex, quadratic regularization term $H^i(\boldPhi^i)$ on Agent $i$'s system response $\boldPhi^i$ to Agent $i$'s cost:
\begin{align} \label{Eqn: tilde J}
    \tilde J^i(\boldz, \boldv, \boldPhi) := J^i(\boldz, \boldv) + H^i(\boldPhi^i).
\end{align}
We now characterize the \textit{best response map}\footnote{The \textit{system response} $\Phi^i$, defined component-wise in Prop. \ref{Prop: SLS, Using Phi to Parameterize K}, and the \textit{best response map} $\BR$, defined in \eqref{Eqn: Best Response Map (BR), Def} describe different mathematical objects.} of each agent $i$, and the set of RCNE nominal trajectories, system responses, and error upper bounds. Below, $\boldz^{-i} \in \R^{(n-n_i)T}$ denotes the nominal trajectories of all agents not indexed $i$, and $\boldv^{-i}$, $\boldPhi^{-i}$, $\boldrho^{-i}$ are defined analogously.

\begin{definition}
% [\textbf{Robustly Constrained Nash Equilibrium}] 
\label{Def: Robustly Constrained Nash Equilibrium (RCNE)}
% \looseness=-1
For each $i \in [N]$, we define Agent $i$'s set of \textit{best responses} given other agents' nominal trajectories, system responses, and error bounds $(\boldz^{-i}, \boldv^{-i}, \boldPhi^{-i}, \boldrho^{-i})$ by:
\begin{align} \label{Eqn: Best Response Map (BR), Def}
    &\BR^i(\boldz^{-i}, \boldv^{-i}, \boldPhi^{-i}, \boldrho^{-i}; \tilde J^i) \\ \nonumber
    := \ &\text{arg} \min_{\boldz^i, \boldv^i, \boldPhi^i, \boldrho^i} \hspace{5mm} \tilde J^i(\boldz, \boldv, \boldPhi) \\ \nonumber
    &\hspace{1cm} \text{s.t.} \hspace{5mm} (\boldz, \boldv, \boldPhi, \boldrho) \in \Safe.
\end{align}
We call a set of agents' nominal state-control trajectories, system responses, and error (upper) bounds $(\boldz^\star, \boldv^\star, \boldPhi^\star, \boldrho^\star)$ a \textit{Robustly Constrained Nash Equilibrium (RCNE)}
% solution of the dynamic game $\G$ 
if, at $(\boldz^\star, \boldv^\star, \boldPhi^\star, \boldrho^\star)$, each agent best responds to all other agents while robustly satisfying all constraints, i.e., for each $i \in [N]$:
\begin{align} \label{Eqn: Robustly Constrained Nash Equilibrium (RCNE), Def}
    \hspace{-3mm}
    (\boldz^{i\star}, \boldv^{i\star}, \boldPhi^{i\star}, \boldrho^{i\star}) \in \BR^i(\boldz^{-i\star}, \boldv^{-i\star}, \boldPhi^{-i\star}, \boldrho^{-i\star}; \tilde J^i).
\end{align}
\end{definition}

% \paragraph{Problem Statement} 
\noindent
\textbf{Problem Statement}:
We aim to 
% compute 
formulate a computationally tractable algorithm for computing all agents' nominal state-control trajectories, system responses, and error upper bounds at Nash equilibrium, as defined in \eqref{Eqn: Robustly Constrained Nash Equilibrium (RCNE), Def}.

\vspace{-1mm}
% \section{Methods}
% \label{sec: Methods}
\section{Algorithm}
\label{sec: Algorithm}

% Although
% % \eqref{Eqn: Best Response Map (BR), Def}-
% \eqref{Eqn: Nash Equilibrium (NE), Def} prescribes a set of optimization problems which characterize the Nash equilibrium solution, \eqref{Eqn: Nash Equilibrium (NE), Def} is not computationally tractable due to 
% the high-dimensional nature

% In Sec. \ref{sec: Methods}
% Below, we will formulate a computationally tractable algorithm for computing the robustly constrained Nash equilibrium.
While a robustly constrained Nash equilibrium can in theory be found by solving \eqref{Eqn: Robustly Constrained Nash Equilibrium (RCNE), Def} for each $i \in [N]$, in practice \eqref{Eqn: Robustly Constrained Nash Equilibrium (RCNE), Def} is computationally intractable to solve with general-purpose solvers, due to the high-dimensional and coupled (across agent indices $i \in [N]$) nature of \eqref{Eqn: Robustly Constrained Nash Equilibrium (RCNE), Def}.
% , stacked across agent indices $i \in [N]$, poses a high-dimensional and coupled optimization problem.
To sidestep these computational difficulties, we present Alg. \ref{Alg: IBR}, which refines the Iterative Best Response (IBR) algorithm \cite[Prop. 1]{Zanardi2021LexicographicGames} in the potential games literature to incrementally search for a Nash equilibrium solution.
% (see Remark ).
% Concretely, Alg. \ref{Alg: IBR} proceeds in $K$ iterations, where $K$, the maximum iteration count, is user-specified. 
Concretely, Alg. \ref{Alg: IBR} proceeds in iterations. Within each iteration, each agent $i \in [N]$ takes its turn applying an incremental best response step, with step size $\alpha \in [0, 1]$, with respect to all other agents' nominal state-control trajectories and system responses (Alg. \ref{Alg: IBR}, Lines \ref{Algline: Compute Best Response}-\ref{Algline: Update Phi, rho via Fast SLS}).
% \footnote{In Alg. \ref{Alg: IBR}, Line \ref{Algline: Increment towards Best Response}, the addition and scalar multiplication operations are executed component-wise.}.

% \looseness-1
It is known that, if each best-response update is performed with maximum step sizes (i.e., $\alpha = 1$ in Alg. \ref{Alg: IBR})
and with zero error, the IBR algorithm converges to a Nash equilibrium of a dynamic potential game \cite{Zanardi2021LexicographicGames}. However, in our robust games formulation, each computation of the best response map \eqref{Eqn: Best Response Map (BR), Def} 
% in Alg. \ref{Alg: IBR}, Line \ref{Algline: Compute Best Response} 
encodes a high-dimensional optimization problem 
% \eqref{Eqn: Best Response Map (BR), Def},
that is difficult to solve directly and efficiently.
% Established results on the convergence of the IBR algorithm, when applied to dynamic potential games, imply that Alg. \ref{Alg: IBR} converges to a Nash equilibrium. We formally record this result below.
Instead, to efficiently approximate the outcome of
Alg. \ref{Alg: IBR}, Line \ref{Algline: Compute Best Response}, we employ the Fast SLS algorithm introduced in \cite{Leeman2024FastSLS}. Crucially, the Fast SLS algorithm exploits the structure stage-wise composition of the overall cost $J^i$ for each agent $i \in [N]$, given by \eqref{Eqn: Overall Cost J i, as sum of Stage-wise costs L ti}, to separately solve for the Agent $i$'s nominal trajectory $(\boldz^i, \boldv^i)$, controller $\K^i$ (through the system response $\boldPhi^i$), and error upper bound $\boldrho^i$. 
The Fast SLS algorithm alternates between updating $K^i$ via a Riccati recursion, optimizing the nominal trajectory $(z^i, v^i)$ as per \eqref{Eqn: Nominal, Dynamics, Agent i}, and performing an additional update of the error upper bound $\boldrho^i$. For details on the original formulation of the fast SLS algorithm, see \cite[Sec. 3]{Leeman2024FastSLS}.
% \antoine{I changed "quadratic programming" into "nominal trajectory". Feel free to revert that change.}
Whereas fast SLS \cite[Sec. 3]{Leeman2024FastSLS} solved QPs to recover solutions to LQR-style problems, our algorithm solves NLPs to recover solutions to more general control problems. 

\begin{remark} \label{Remark: Compare Fast SLS on nominal problem}
\looseness-1To motivate our use of the IBR algorithm, we note that, when applied to $N$-agent, $T$-horizon dynamic games in which the average agent state and control dimensions are $\bar n$ and $\bar m$, respectively, our Fast SLS and IBR-based approach yields a 
% . Each best response step alternates a Riccati-style feedback update with a trajectory update.
% yielding 
per-iteration complexity $\mathcal{O}(T^2 N (\bar n + \bar m)^3)$. (Here, \say{per-iteration complexity} is associated with the time needed to execute Alg. \ref{Alg: IBR}, Lines \ref{Algline: Compute Best Response}-\ref{Algline: Update Phi, rho via Fast SLS}). In contrast, a centralized implementation of the form \cite[Alg. 1]{Kavuncu2021PotentialiLQR} would yield a per-iteration complexity of $\mathcal{O}(T^2 N^3 (\bar n + \bar m)^3)$ to update each agent's trajectory and control designs, even when accelerated via Fast SLS.
\vspace{-7pt}
\end{remark} 

\begin{algorithm}[ht] 
\caption{Iterative Best Response (IBR) 
% \frank{Version 2} 
for Computing RCNE}
\label{Alg: IBR} 
\KwIn{Maximum iteration count $K$, Error Tolerance $\epsilon$, Step size $\alpha \in [0, 1]$, Augmented Costs $\{\tilde J^i(\cdot): i \in [N] \}$, Initial nominal trajectories and system responses $\{(\boldz^{i,(0)}, \boldv^{i,(0)}, \boldPhi^{i,(0)}, \boldrho^{i,(0)}): i \in [N] \}$} 
\For{$k = 1, \cdots, K$}{ 
    \For{$i = 1, \cdots, N$}{ 
        $(\hat\boldz^{i,(k)}, \hat\boldv^{i,(k)}, \hat\boldPhi^{i,(k)}, \hat\boldrho^{i,(k)}) \in \BR^i(\boldz^{-i,(k)}, \boldv^{-i,(k)}, \boldPhi^{-i,(k)}, \boldrho^{-i,(k)}; \tilde J^i)$
        \nllabel{Algline: Compute Best Response}

        % Update $(\boldz^{i,(k)}, \boldv^{i,(k)}, \boldPhi^{i,(k)}, \boldrho^{i,(k)}) \gets (1-\alpha) (\hat\boldz^{i,(k)}, \hat\boldv^{i,(k)}, \hat\boldPhi^{i,(k)}, \hat\boldrho^{i,(k)}) + \alpha (\boldz^{i,(k)}, \boldv^{i,(k)}, \boldPhi^{i,(k)}, \boldrho^{i,(k)})$
        % \nllabel{Algline: Increment towards Best Response}

        % Update 
        $\boldv^{i,(k)} \gets \alpha \hat\boldv^{i,(k)} + (1-\alpha) \boldv^{i,(k)}$
        \nllabel{Algline: Increment v towards Best Response}

        % Update 
        $\boldz^{i,(k)} \gets$ Unroll \eqref{Eqn: Nominal, Dynamics, Agent i} using $\boldv^{i,(k)}$ \nllabel{Algline: Update z via nominal dynamics}

        $\boldPhi^{i,(k)}, \boldrho^{i,(k)} \gets$ Implement Riccati updates in Fast SLS using $\boldz^{i,(k)}$, $\boldv^{i,(k)}$. \nllabel{Algline: Update Phi, rho via Fast SLS}

    } 
    
    $\delta^{(k)} \gets \max_{i \in [N]} \max_{\substack{t,\tau \in [T] \\ t \geq \tau}} \max \big\{ \Vert \boldz^{i,(k)} - \boldz^{i,(k-1)} \Vert_2, \Vert \boldv^{i,(k)} - \boldv^{i,(k-1)} \Vert_2, \Vert \boldPhi_{t,\tau}^{i,(k)} - \boldPhi_{t,\tau}^{i,(k-1)} \Vert_2, \Vert \boldrho^{i,(k)} - \boldrho^{i,(k-1)} \Vert_2 \big\}$.
    
    \eIf{$\delta^{(k)} \leq \epsilon$}{
        $(\boldz^\star, \boldv^\star, \boldPhi^\star, \boldrho^\star) \gets (\boldz^{(k)}, \boldv^{(k)}, \boldPhi^{(k)}, \boldrho^{(k)})$.
        
        Break
    }
    {
        $(\boldz^{(k+1)}, \boldv^{(k+1)}, \boldPhi^{(k+1)}, \boldrho^{(k+1)}) \gets (\boldz^{(k)}, \boldv^{(k)}, \boldPhi^{(k)}, \boldrho^{(k)})$.
    }
} 
\KwOut{$\boldz^\star, \boldv^\star, \boldPhi^\star, \boldrho^\star$}
\end{algorithm} 
% \vspace{-1mm}

% \begin{algorithm}[ht] 
% \caption{Iterative Best Response (IBR) for Computing RCNE \frank{Version 3} } 
% % \label{Alg: IBR} 
% \KwIn{Augmented Costs $\{\tilde J^i(\cdot): i \in [N] \}$, Initial nominal trajectories and system responses $\{(\boldz^{i,(0)}, \boldv^{i,(0)}, \boldPhi^{i,(0)}): i \in [N] \}$} 
% \While{\text{not converged}}{ 
%     \For{$i = 1, \cdots, N$}{ 
%     % \nllabel{Algline: Best Response}
%         Update $(\boldz^{i,(k)}, \boldv^{i,(k)}, \boldPhi^{i,(k)}) \in \BR^i(\boldz^{-i,(k)}, \boldv^{-i,(k)}, \boldPhi^{-i,(k)}; \tilde J^i)$
%     } 
    
%     \eIf{$(\boldz^{(k)}, \boldv^{(k)}, \boldPhi^{(k)}) = (\boldz^{(k-1)}, \boldv^{(k-1)}, \boldPhi^{(k-1)})$}{
%         $(\boldz^\star, \boldv^\star, \boldPhi^\star) \gets (\boldz^{(k)}, \boldv^{(k)}, \boldPhi^{(k)})$.
        
%         Break
%     }
%     {
%         $(\boldz^{(k+1)}, \boldv^{(k+1)}, \boldPhi^{(k+1)}) \gets (\boldz^{(k)}, \boldv^{(k)}, \boldPhi^{(k)})$.
%     }
% } 
% \KwOut{$\boldz^\star, \boldv^\star, \boldPhi^\star$}
% \end{algorithm} 

% \vspace{-35pt}
\section{Experiments}
\label{sec: Experiments}

% By evaluating our method across a broad range of simulation and hardware experiments, 

We evaluated our algorithm by generating interactive, robustly constraint-satisfying motion plans across a broad range of multi-agent interaction scenarios in simulation and on hardware. First, we present experiment settings used in our reported experiments, such as agent costs and constraint types (Sec. \ref{subsec: Experiment Setup}). 
We then evaluate our method on a \say{narrow corridor} scenario, wherein trajectories generated by our method remain robustly safe in confined spaces despite dynamics noise (Sec. \ref{subsec: Narrow Corridor Experiment}). Our approach efficiently generates robustly constraint-satisfying motion plans in scenarios with up to 24 4D-unicycle agents (Sec. \ref{subsec: Scaling Experiment}), with state-dependent dynamics noise (Sec. \ref{subsec: State-Dependent Noise Experiment}), or heterogeneous robot teams of 
% 1 quadcopter and 2 4D-unicycle ground vehicles each
quadcopters and 4D-unicycle ground vehicles
(Sec. \ref{subsec: Heterogeneous Team Simulation}, \ref{subsec: Heterogeneous Team Hardware Experiment}). In contrast, while operating under
baseline open-loop game-theoretic motion plans \cite{lecleach2020algames}, the same multi-robot systems often violated collision-avoidance constraints when deployed in narrow spaces or in the presence of state-dependent dynamics noise (Sec. \ref{subsec: Narrow Corridor Experiment} and \ref{subsec: State-Dependent Noise Experiment}). 
Our codebase can be found at: \url{https://github.com/nexuszhan/Robust-Dynamic-Game-SLS}.

\vspace{-2mm}
\subsection{Experiment Setup}
\label{subsec: Experiment Setup}

% \looseness-1
Our experiments in Sec. \ref{subsec: Narrow Corridor Experiment}-\ref{subsec: Heterogeneous Team Hardware Experiment} use the following types of agent dynamics, constraints, and costs. Below, given the state $x_t^i$ of an agent $i$ at time $t$, we denote by $p_t^i$, $p_{x,t}^i$, $p_{y,t}^i$, and $p_{z,t}^i$ the component(s) of $x_t^i$ corresponding to the position vector, $x$-position, $y$-position, and $z$-position respectively.

\paragraph{Dynamics models}
\looseness-1We apply our methods to plan trajectories for ground robots with 4D unicycle and airborne robots with 12D quadcopter dynamics in simulation (Sec. \ref{subsec: Narrow Corridor Experiment}-\ref{subsec: Heterogeneous Team Simulation}), as well as ground robots with 3D Dubins car dynamics and airborne robots with 3D single integrator dynamics on hardware (Sec. \ref{subsec: Heterogeneous Team Hardware Experiment}).
Across experiments, unless otherwise stated, we discretize the above continuous-time dynamics models at intervals of $\Delta t = 0.1$ and use a time horizon of $T = 80$. 
% In alignment with the dynamics model presented in Sec. \ref{sec: Preliminaries}, 
All noise terms $w_t^i$ are drawn from the unit $2$-norm ball $\Ball_2^{n_i}$. We set $E_t^i(x_t^i) = 0.002 \ I_4$ unless otherwise stated (e.g., in Sec. \ref{subsec: State-Dependent Noise Experiment}, we define $E_t^i(x_t^i)$ as state-varying.)
% Sec. \ref{subsec: Narrow Corridor Experiment} and \ref{subsec: Scaling Experiment}, we set $E_t^i(x_t^i) = 0.002 \ I_4$, while in Sec. \ref{subsec: State-Dependent Noise Experiment} and Sec. \ref{subsec: Heterogeneous Team Hardware Experiment}, we define $E_t^i(x_t^i)$ to be state-varying. 

\paragraph{Constraint types}
We consider \textit{agent-specific} constraints that encode component-wise bounds on position, orientation, linear velocity, angular velocity, and linear acceleration. 
% For instance, given the state $x_t^i$ of an agent $i$ at time $t$, component-wise boundary constraints on its position vector $p_t^i$ with designated boundaries $\hat p_{t,i}^i$ are encoded by constraints of the form $g_{t,k}^i(x_t^i) := p_t^i - \hat p_t^i \leq 0$ or $g_{t,k}^i(x_t^i) := - p_t^i - \hat p_t^i \leq 0$, with the inequality given coordinate-wise. 
% Collision avoidance, at each time $t$ between two agents $i, j \in [N]$ with radii of $r^i$ and $r^j$, are given by shared constraints of the form $h_{t,k}(x_t) := - \Vert p_t^i - p_t^j \Vert_2 + r^i + r^j \leq 0$.
We also implement \textit{shared} constraints $h_{t,k}$ between pairs of agents $i, j \in [N]$ (say, with radii of $r^i$ and $r^j$), that encode (i) collision-avoidance, by setting $h_{t,k}(x_t) := - \Vert p_t^i - p_t^j \Vert_2 + r^i + r^j \leq 0$; (ii) proximity constraints, by setting $h_{t,k}(x_t) := \Vert p_t^i - p_t^j \Vert_2 - \hat r^{ij} \leq 0$, for some prescribed proximity distance $r^{ij} > r^i + r^j$; and (iii) line-of-sight constraints, of form described in \cite[App. D.3]{Zhang2025ConstraintLearninginMultiAgentDynamicGames}. Each line-of-sight constraint enforced on a pair of robots, say, $i$ and $j$, requires the angle between (i) velocity vector of the first agent, and (ii) the relative position of the second agent to the first, to always be within a prescribed bound $[-\theta^{ij}, \theta^{ij}]$.

\paragraph{Agent Costs}
Across our experiments, the cost $J^i$ of each agent $i$ will be a sum of some of the cost components $C_1^i, C_2^i, C_3^i, C_4^i: \R^{(n_i+m_i)T} \ra \R$ defined below, which encode an LQR cost, a smoothness and goal-reaching cost, a collision avoidance cost, and a proximity cost, respectively:\footnote{In many of our experiments, we encode collision avoidance using both a \textit{hard} constraint and a \textit{soft} penalty.}
\begin{align} 
\label{Eqn: LQR trajectory cost}
    \textstyle C_1^i(\boldx, \boldu) &= ({x_T^i} - \hat x_T^i)^\top Q_f^i ({x_T^i} - \hat x_T^i) \\ \nonumber
    &\hspace{1mm} + \textstyle\sum_{t=0}^{T-1} \big[({x_t^i}-\hat x_T^i)^\top Q^i (x_t^i-\hat x_T^i) + {u_t^i}^\top R^i u_t^i) \big], \\ \label{Eqn: Smoothness trajectory cost}
    % &C_2^i(\boldx, \boldu) \\ \nonumber
    \textstyle C_2^i(\boldx, \boldu) &= ({x_T^i} - \hat x_T^i)^\top Q_f^i ({x_T^i} - \hat x_T^i) \\ \nonumber
    &\hspace{1mm} + \textstyle \sum_{t=1}^{T} (x_{t}^i-x_{t-1}^i)^\top Q^i (x_t^i-x_{t-1}^i), \\
    \label{Eqn: Coupled collision avoidance cost}
    \textstyle C_3^{ij}(\boldx, \boldu) &= - 0.001 \cdot \textstyle \sum_{t=0}^T \Vert p_t^i - p_t^j \Vert_2^2, \\ \label{Eqn: Coupled proximity cost}
    \textstyle C_4^{ij}(\boldx, \boldu) &= 0.001 \cdot \textstyle \sum_{t=0}^T \Vert p_t^i - p_t^j \Vert_2^2
\end{align}
Above, for each agent $i$, $\hat x_T^i \in \R^{n_i}$ denote specified goal states, while $Q_f^i, Q^i \in \R^{n_i \times n_i}$ (resp., $R^i \in \R^{m_i \times m_i}$) are symmetric positive semi-definite (resp., definite) weight matrices. The above parameters in general vary from one experiment to another, and will be specified as appropriate.

% In each simulation, 
Unless specified elsewise, we set $J^i(\boldx, \boldu) = C_2^i(\boldx, \boldu) + \sum_{j \ne i} C_3^{ij}(\boldx, \boldu)$ when evaluating our method, \revision{which renders the resulting game a dynamic potential game}. We generate baseline results using ALGAMES with the same cost $J^i$.
% $J^i(\boldx, \boldu) = C_2^i(\boldx, \boldu) + C_3^{ij}(\boldx, \boldu)$.

% \paragraph{Algorithm Parameters}
\paragraph{Algorithm Implementation}
For the parameters used in Alg. \ref{Alg: IBR}, we fixed a maximum iteration count of $K = 5$ and an error tolerance of $\epsilon = 0.001$. Although we varied the step size $\alpha$ across experiments (see Secs. \ref{subsec: Narrow Corridor Experiment}-\ref{subsec: Heterogeneous Team Hardware Experiment} below) for the first $Q-1$ iterations, we fixed $\alpha = 1$ for the last iteration to enforce robust constraint satisfaction. Our experimental results indicate that, even with a limited number of iterations, Alg. \ref{Alg: IBR} generates solutions that encode reasonable multi-agent interactions while always satisfying constraints robustly.
We run all experiments on an Intel Core i9-13950HX CPU.

\paragraph{Hessian bounds and Lipschitz constants}
To reduce the conservativeness of the robust controller generated by Alg. \ref{Alg: IBR}, we adopt a simplified implementation by setting the Hessian bounds and Lipschitz constants $\mu_{t,k}^i$, $\chi_{t,k}^i$, $\psi_{t,k}^i$, and $L_{E,t,k}^i$ to zero in \eqref{Eqn: f tki Hessian bound, denoted mu tki}-\eqref{Eqn: L etk i}. \revision{Alternatively, data-driven estimates of these constants can be obtained to provide less conservative, high-probability guarantees by following \cite{srinivasan2026safety}.} Although setting these constants to zero implies that we are not guaranteed to over-approximate the error $e_t^i$ between the nominal and realized trajectories, across our experiments, our algorithm still \textit{empirically} satisfy all constraints robustly.

\paragraph{Figure Conventions}
In each figure depicting trajectory rollouts from by our method, circles or balls indicate the Minkowski sum of the volume of each agent, as given by its radius $r^i$, and regions of realizable trajectories, as computed from the error upper bound values $\boldrho^i$. Figures of trajectories generated by the ALGAMES baseline \cite{lecleach2020algames} depict colored circles that indicate merely the volume of each agent.

% ~\\
% \cyc{To list in paragraph environment:
% \begin{enumerate}
%     \item Dynamics types---4D unicycle, 12D quadcopter \cite{sabatino2015PhDThesisquadrotor}, 3D single integrator, and 3D Dubins car.
%     \item Constraint types---(1) Individual constraints: component-wise bounds, on position, linear velocity, orientation, linear velocity, angular velocity, and linear acceleration vectors; (2) Team constraints: collision avoidance bounds.
%     \item Agent Costs---(1) LQR; (2) Goal reaching and smoothness; (3) Collision avoidance costs; add footnote explaining to the reviewer that collision avoidance is both enforced as a (hard) constraint and penalized as a (soft) cost.
%     \item (Skip on first pass) Disturbance rejection costs $H$---As reported in the params.xlsx file.
%     \item Other Parameters---?
%     \item Figure conventions---Circles/balls: feasible state realizations as computed from error upper bounds $\boldrho^i$.
% \end{enumerate}
% }
% ~\\

\subsection{Narrow Corridor Experiment}
\label{subsec: Narrow Corridor Experiment}

\begin{figure}
    \centering
    \includegraphics[width=\linewidth]{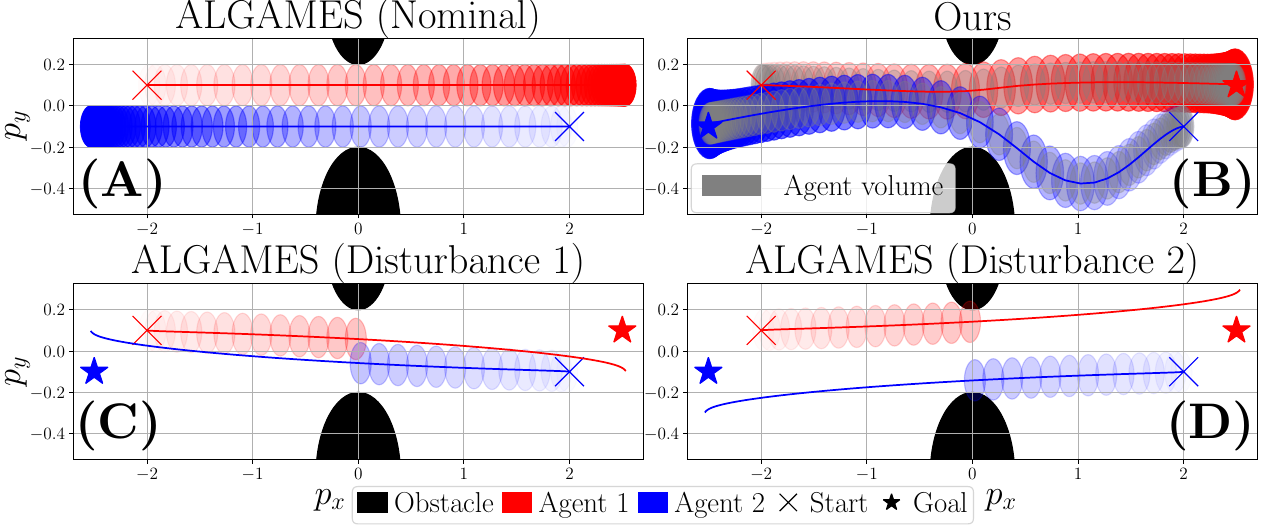}
    \vspace{-15pt}
    \caption{
    \textbf{Narrow Corridor Experiment}: 
    In a confined space, two 4D unicycle agents attempt to bypass each other safely despite dynamics noise.  
    Across 500 rollouts, our method always generates safe trajectories (B), while the ALGAMES baseline \cite{lecleach2020algames} violated constraints 91.8\% of the time.
    % 
    % The figure shows the narrow corridor experiment with two agents, each modeled as a 4D unicycle with radius 0.1 m. The environment contains two circular obstacles at (0, 0.5) and (0, –0.6) with radii 0.3 m and 0.4 m, respectively. Both methods plan for 60 time steps (0.1s each).
    % The trajectory cost function of our method is combination of \eqref{Eqn: LQR trajectory cost} and \eqref{Eqn: Coupled collision avoidance cost}. 
    % ALGAMES assumes nominal dynamics and produces straight nominal trajectories, whereas our method accounts for state-dependent noise (modeled as bounded additive noise scaled by the state) and shows agents evading each other while staying within robust reachable sets (visualized as colored circles).
    }
    \label{fig:narrow_corridor}
    \vspace{-10pt}
\end{figure}

\looseness-1First, we generate trajectories in a \textit{narrow corridor} experiment using our Alg. \ref{Alg: IBR} and the baseline ALGAMES approach \cite{lecleach2020algames} (Fig. \ref{fig:narrow_corridor}). In this experiment, 2 ground robots, each with radius $0.1$m, attempt to bypass each other in a confined space over $T = 60$ time steps while robustly satisfying the set of prescribed constraints described below, despite dynamics noise. 
% Each agent's $x$-$y$ coordinates, orientation (counterclockwise angle with respect to the $+x$-axis), linear velocity, angular velocity, and linear acceleration are constrained to stay within the bounds $[-100, 100]^2$, $[-10, 10]$, $[-1, 2.2]$, $[-\pi/2, \pi/2]$, and $[-1, 1]$, respectively. 
Each agent must satisfy collision-avoidance constraints with respect to all other agents, as well as two obstacles, one at $(0, 0.5)$ with radius $0.3$m and one at $(0, -0.6)$ with radius $0.4$m. 
% When implementing our Alg. \ref{Alg: IBR}, we 
For our Alg. \ref{Alg: IBR} implementation, we set $\alpha = 0.3$, and use the cost $J^i(\boldx, \boldu) = C_1^i(\boldx, \boldu) + \sum_{j \ne i} C_3^{ij}(\boldx, \boldu)$, with $Q = 2I_4$, $R = I_2$, and $Q_f = \diag\{5, 5, 0, 5\}$. To implement the ALGAMES baseline, we set $Q = I_4$, $R = I_2$, and $Q_f = I_4$.

% As illustrated in Fig. \ref{fig:narrow_corridor}, 
Across 500 rollouts, our method consistently generates interactive trajectories which enable the two agents to safely bypass each other while satisfying all of the constraints above (Fig. \ref{fig:narrow_corridor}B), despite dynamics noise. Our Alg. \ref{Alg: IBR} runtime was 32.2s.
In contrast, out of 500 rollouts with dynamics noise, the baseline ALGAMES algorithm plans trajectories which resulted in collisions 91.8\% of the time (Fig. \ref{fig:narrow_corridor}A, C, D).

% We compare our method with ALGAMES in a scenario where there is a narrow corridor that can exactly contain two agents in parallel as shown in Fig. \ref{fig:narrow_corridor}. Each agent is a 4D unicycle with a radius of 0.1m and two obstacles are located at (0, 0.5) and (0, -0.6) with radius 0.3m and 0.4m respectively. Both methods plan for 60 time steps where each time step is 0.1s. Since ALGAMES assumes nominal dynamic system, it plans two straight trajectories. However, the straight trajectories are prone to cause collisions when there is disturbance in the system with 87 collisions in 100 rollouts with its open-loop controller. In contrast, our method enables two agents to evade each other in the corridor (runtime: 32.2s). The average difference between goal position and actual terminal position per agent is 0.0366m for ALGAMES and 0.0169m for our method. 
% Upper part of Fig. \ref{fig:narrow_corridor} shows planned trajectories of ALGAMES and our method. ALGAMES assumes nominal dynamics and produces straight nominal trajectories, whereas our method accounts for disturbances ($E=2e^{-3}I_4$) and shows agents evading each other while staying within robust reachable sets (visualized as colored circles). Lower part shows that collisions happen when we apply ALGAMES's inputs and the collision rate is $91.8\%$ out of 500 rollouts when norms of disturbances are 1. In contrast, our method achieves collision-free performance under the same disturbances using the feedback controller synthesized using Alg. \ref{Alg: IBR}. 

\subsection{Scaling Experiment}
\label{subsec: Scaling Experiment}

\begin{figure}
    \centering
    \includegraphics[width=0.99\linewidth]{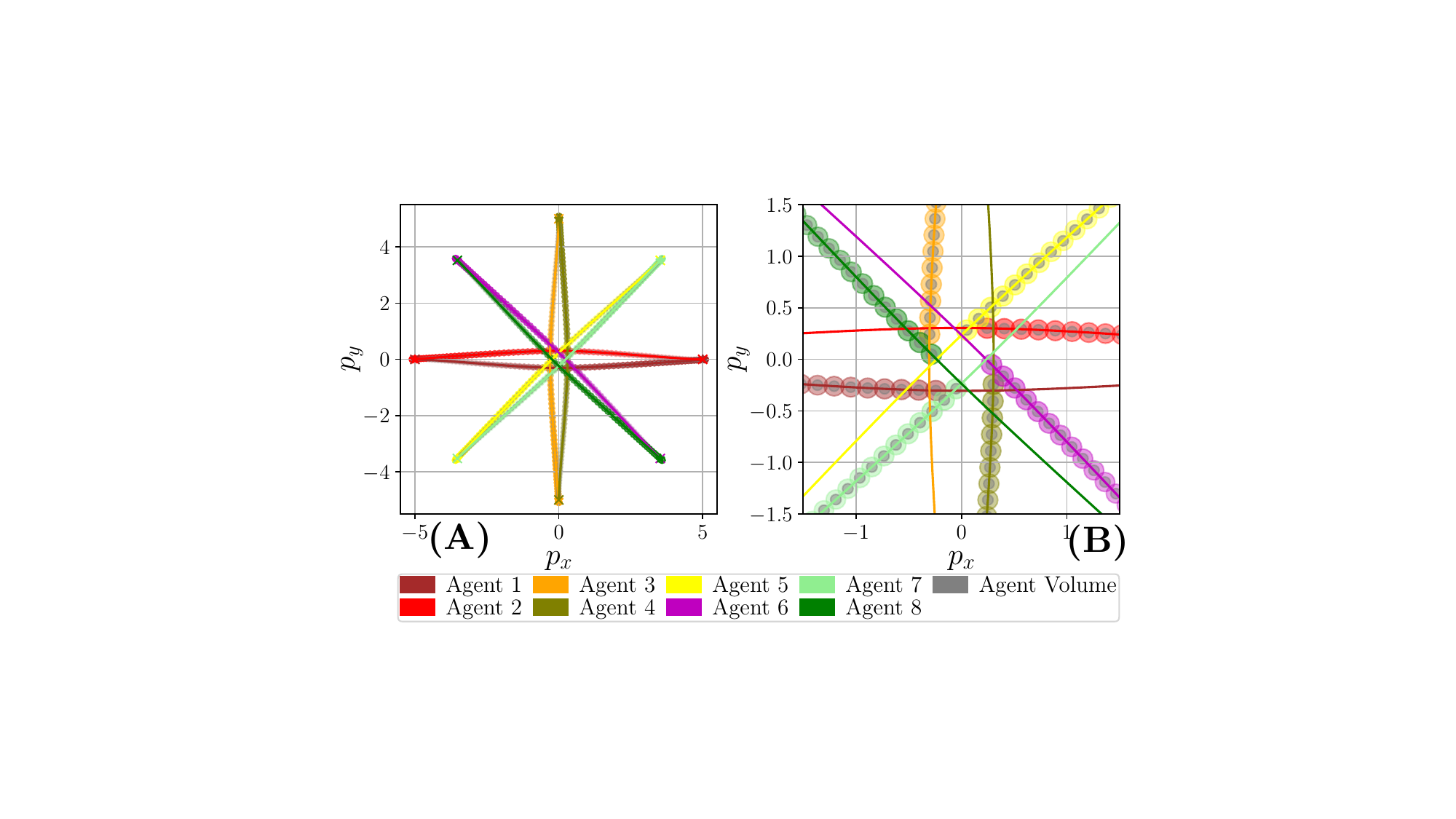}
\caption{\textbf{Scaling Example}: (A) Our method synthesizes robustly collision-free, interactive trajectories for 8 4D-unicycle agents. (B) presents a snapshot of (A) at $t = 40$ when the agents are in close proximity to each other.
% Trajectories planned for $8$ agents. 
% Each agent is modeled as a 4D unicycle, and the task is to exchange positions across the workspace over $80$ time steps ($0.1\,\text{s}$ per step). 
% Coupled costs enforce collision avoidance and coordination among agents. 
% The left figure shows how our method synthesizes robust, collision-free and cooperative trajectories while respecting interaction constraints between multiple agents.
% The right figure shows the trajectories at time step 40 when agents become close.
}
    \label{fig:scaling}
    \vspace{-13pt}
\end{figure}

Next, we demonstrate that our method efficiently generates robustly constraint-satisfying trajectories even for multi-agent games with large numbers of agents.
Concretely, for $N = 4, 8, 16$ and $24$, we computed robust motion plans for $N$ 4D-unicycle agents, each with radius 0.05m, who aim to navigate while robustly satisfying pre-specified constraints despite dynamics noise (See Fig. \ref{fig:scaling} for the $N = 8$ setting). 
% Each agent's $x$-$y$ coordinates, orientation, linear velocity, angular velocity, and linear acceleration are constrained to stay within the bounds $[-100, 100]^2$, $[-100, 100]$, $[-3, 3]$, $[-\pi/2, \pi/2]$, and $[-3, 3]$, respectively. 
Each agent must also satisfy collision-avoidance constraints with respect to all other agents.
% Moreover, all agents must avoid two obstacles, one at $(0, 0.5)$ with radius $0.3$m and one at $(0, -0.6)$ with radius $0.4$m. 
% When implementing our Alg. \ref{Alg: IBR}, we 
We 
% use the learning rate
set $\alpha = 0.1$ for the $N = 4, 8$ settings and $\alpha = 0.5$ for the $N = 16, 24$ settings. 
We also set $Q = 2I_4$, $R = O$, and $Q_f = \diag\{10, 10, 0, 10\}$. Our Alg. \ref{Alg: IBR} runtimes for the $N = 4, 8, 16, 24$ settings are respectively given by 68.3s, 153.4s, 462.9s and 891.1s, with standard deviations 0.7s, 0.3s, 4.2s and 5.9s. We generated 10 trajectory rollouts per agent for each of the $N = 4, 8, 16, 24$ settings, all of which robustly satisfied all constraints.

% To show that our method can deal with large number of agents, we let 4, 8, 16 and 24 agents exchange positions with opposite agent. The agents are modeled as 4D unicycle with radius of 0.05m and we plan for 80 time steps with each time step be 0.1s. The average IBR runtimes are 68.3s, 153.4s, 462.9s and 891.1s with standard deviations 0.7s, 0.3s, 4.2s and 5.9s respectively. 

% \begin{table}[h]
%     \centering
%     \begin{tabular}{|c|c|c|c|}
%         \hline
%         N & Runtime & N & Runtime \\
%         \hline
%         4 & 1m4.4s & 8 & 3m12.5s \\
%         \hline
%         16 & 7m43.8s & 24 & 14m41.2s \\
%         \hline
%     \end{tabular}
%     \caption{Runtime for 4, 8, 16 and 24 agents.}
% \end{table}
% \begin{figure}
%     \centering
%     \includegraphics[width=0.9\linewidth]{img/scaling_runtime.pdf}
% \caption{Runtime scaling with the number of agents. 
% We evaluate scenarios with $4$, $8$, $16$, and $24$ agents, each modeled as a 4D unicycle with radius $0.05\,\text{m}$. 
% Agents are tasked with exchanging positions with their opposite counterparts over $80$ time steps ($0.1\,\text{s}$ per step). 
% The plot shows runtime growth as the number of agents increases, demonstrating the scalability of our method to high-dimensional multi-agent systems.}

%     \label{fig:scaling_runtime}
% \end{figure}

\subsection{State-Dependent Noise Experiment}
\label{subsec: State-Dependent Noise Experiment}

\looseness-1We show that Alg. \ref{Alg: IBR} produces interactive trajectories that robustly satisfy constraints in multi-agent interactions with \textit{state-dependent} noise, i.e., when $E_t^i(\cdot)$ is \textit{not} constant, while ALGAMES generates constraint-violating motion plans. Specifically, we consider 4 ground robots who wish to navigate a shared intersection while robustly satisfying constraints, located at the origin of a given 2D coordinate system, despite state-dependent dynamics noise (Fig. \ref{fig:state-dependent-noise}). In particular, we design $E_t^i(x_t^i)$ as shown below, to promote large realizations of dynamics noise near the origin, where agents would be in close proximity near the middle of the time horizon:
\begin{align} \label{Eqn: Exp, E, Gaussian PDF-like}
    E_t^i(x_t^i) = \textstyle\frac{1}{1000\pi} e^{-25(p_{x,t}^2 + p_{y,t}^2)} I_4.
\end{align}
% Each agent's $x$-$y$ coordinates, orientation, linear velocity, angular velocity, and linear acceleration are constrained to stay within the bounds $[-5, 5]^2$, $[-10, 10]$, $[-0.1, 0.2]$, $[-\pi/2, \pi/2]$, and $[-0.1, 0.1]$, respectively. 
Each agent must also satisfy collision-avoidance constraints with respect to all other agents.
When implementing our Alg. 1, 
% we set $\alpha = 1.0$ for the $N = 4, 8$ settings and $\alpha = 0.5$ for the $N = 16, 24$ settings, and we use
% we use the cost $J^i(\boldx, \boldu) = C_2^i(\boldx, \boldu) + C_3^{ij}(\boldx, \boldu)$. 
we set $\alpha = 1$, and choose $Q = 2I_4$, $R = O$, and $Q_f = \diag\{10, 10, 0, 10\}$. 
For our ALGAMES baseline, we choose $Q = \diag\{3, 3, 0, 1\}$, $R = O_{4 \times 4}$, and $Q_f = \diag\{3, 3, 0, 1\}$. 

Across 100 trajectory rollouts, our method consistently produced constraint-satisfying plans despite state-dependent dynamics noise (Fig. \ref{fig:state-dependent-noise}), while enabling each agent to reach an average distance of 4.04cm away from their goal position. Our Alg. \ref{Alg: IBR} runtime was 60s. In contrast, all 100 trajectory rollouts produced by the ALGAMES baseline violated the constraints, with average goal deviation of 9.82cm.
\vspace{-1mm}
% Fig. \ref{fig:state-dependent-noise}

% We show the advantage of integrating nominal trajectory planning with robust control in Fig. \ref{fig:state-dependent-noise}. Our method enable agents to traverse low-disturbance area instead of choosing shorter trajectories exposed to higher disturbance. ALGAMES has 100\% collision rate and 0.0982m deviation from goal positions per agent in this environment compared to 0\% collision rate and 0.0404m deviation of our method when disturbances with norm 1 are sampled out of 100 rollouts.
% which enables agents to traverse low-disturbance area instead of choosing shorter trajectories but being exposed to higher disturbance. We simulate an environment such that the disturbance follows a 2D Gaussian distribution $E(px,py)=\frac{1}{2\pi} e^{-0.5(px^2+py^2)/0.02}$ around the center. Each agent is a 4D unicycle with a radius of 0.06m and both methods plan for 80 time steps where each time step is 0.1s. As shown in Fig. \ref{fig:state-dependent-noise}, our method plans trajectories that get around the area with poor traversibility. ALGAMES has 100\% collision rate and 0.0982m deviation from goal positions per agent in this environment compared to 0\% collision rate and 0.0404m deviation of our method. 

\begin{figure}
    \centering
    \includegraphics[width=\linewidth]{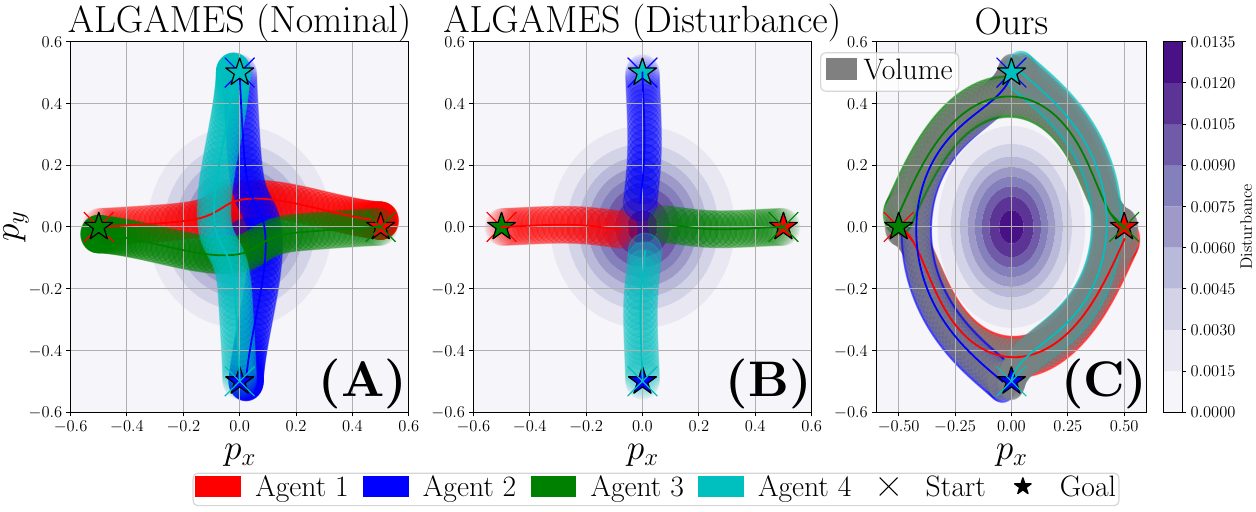}
\caption{\textbf{State-Dependent Noise Example}:
Four 4D-unicycle agents aim to cross an intersection at the same time despite state-dependent noise with maximum noise realization at the intersection (see \eqref{Eqn: Exp, E, Gaussian PDF-like}). Across 100 rollouts, our method always generates safe trajectories (C), while the ALGAMES baseline \cite{lecleach2020algames} generated trajectories that always violated the constraints, in both the noise-free (A) and noisy (B) settings.
% 
% , each modeled as a 4D unicycle with radius $0.06\,\text{m}$, are tasked to exchange positions over $80$ time steps ($0.1\,\text{s}$ per step). 
% Disturbances follow a 2D Gaussian distribution 
% $E(p_x, p_y) = \tfrac{0.002}{2\pi} e^{-0.5(p_x^2 + p_y^2)/0.02} I_4$ 
% centered at the origin where $p_x$ and $p_y$ are $x$ and $y$ positions, creating a high-uncertainty region. 
% ALGAMES, which assumes nominal dynamics does not account for the uncertainties, in contrast to our method which produce safer detours around the disturbance area, achieving $0\%$ collisions and an average deviation of $0.0404\,\text{m}$.
}

    \label{fig:state-dependent-noise}
    \vspace{-18pt}
\end{figure}

\vspace{-1mm}
\subsection{Heterogeneous Team Simulation}
\label{subsec: Heterogeneous Team Simulation}
\begin{figure}
    \centering
    \includegraphics[width=0.95\linewidth]{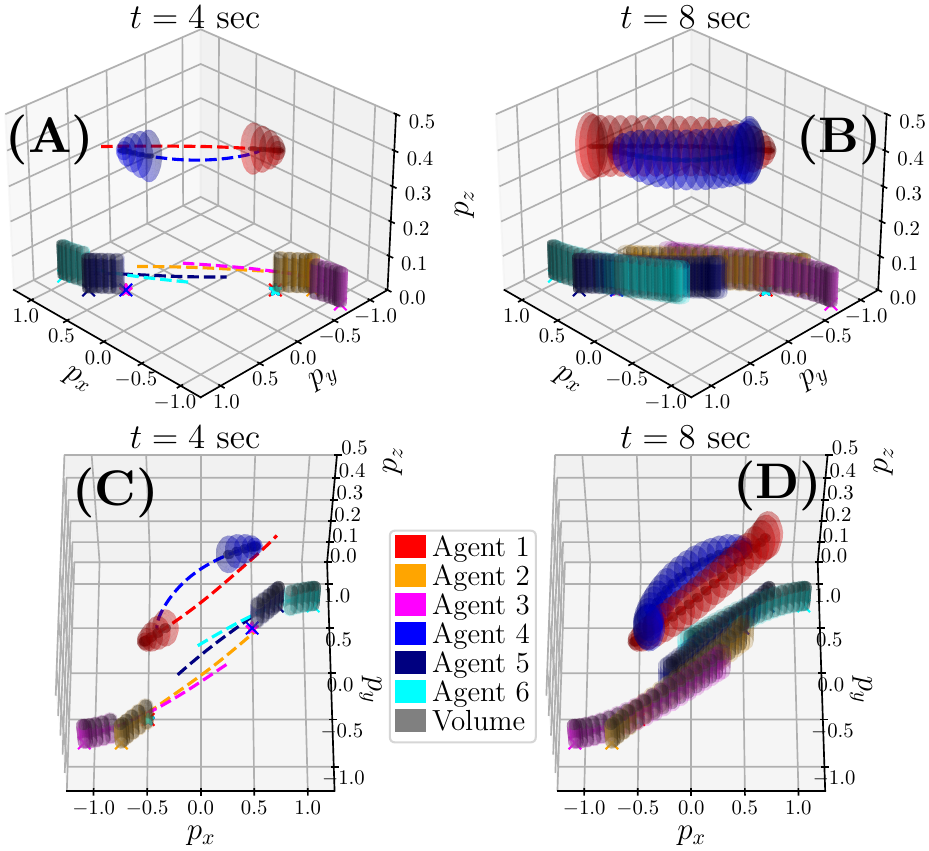}
    \vspace{-8pt}
\caption{\textbf{Heterogeneous Team Simulation}: 
% Simulation of two interacting heterogeneous teams. 
In each of two teams, one ground robot follows a second, which in turn follows a quadcopter. 
% Two teams are formed, each consisting of one quadcopter leader and two unicycles. 
% The quadcopters are modeled as $12$D systems with radius $0.05\,\text{m}$, while the unicycles are modeled as $4$D systems with radius $0.06\,\text{m}$. In each team, one unicycle follows the quadcopter while another follows quadcopter's folloer. 
Each team must satisfy proximity and line-of-sight constraints besides collision avoidance with other agents. 
Our Alg. \ref{Alg: IBR} produced trajectories satisfying all pre-specified proximity, line-of-sight, and collision avoidance constraints despite dynamics noise.
% Followers must stay within $0.5\,\text{m}$ of their leaders under a $90^{\circ}$ field-of-view. 
% Quadcopters must stay within $0.5\,\text{m}$ of their followers under a $90^{\circ}$ field-of-view, and unicycles must maintain proximity and line-of-sight with their leaders. 
% Trajectories are planned for $80$ time steps ($0.1\,\text{s}$ per step), with results plotted every $4$ steps. 
% The IBR runtime is 5m23s.
(A), (C) (resp., (B), (D)) show trajectories from two angles at $t=4$s (resp., $t = 8$s). 
% and (B), (D) shows trajectories at $t=8$s from 2 angles. 
% The figure shows that our method generates robust trajectories satisfying proximity, visibility, and collision-avoidance constraints simultaneously.
}
    \label{fig:heter_team_sim}
    \vspace{-19pt}
\end{figure}

% \vspace{-2mm}
\looseness-1In Sec. \ref{subsec: Heterogeneous Team Simulation}-\ref{subsec: Heterogeneous Team Hardware Experiment}, we show Alg. \ref{Alg: IBR} computes robustly constraint-satisfying interactive motion plans for heterogeneous robot teams in simulation and on hardware. We first simulate two teams of three robots: a 4D unicycle ground robot follows a second, which follows a 12D quadcopter. Each ground robot and quadcopter has radius $r^i = 0.06$m and $r^i = 0.05$m, respectively.
We set $E_t^i(x_t^i) = \diag\{ 0.004 \times I_3, 1 \times 10^{-5} I_9 \}$ 
for each quadcopter, where the first 3 coordinates represent the quadcopters' $x,y,z$ coordinates, and $E_t^i(x_t^i) = 5 \times 10^{-4} I_4$ for each ground robot.
Each pair of robots 
% $i, j \in [N]$ 
must satisfy collision-avoidance constraints. Each Dubin car must also satisfy, with respect to its leader, proximity constraints with distance $\rho^{ij} = 0.5$m and line-of-sight constraints with maximum angle $\theta^{ij} = \pi/2$ rad.
When running Alg. \ref{Alg: IBR}, 
we set $\alpha = 0.2$ and $J^i(\boldx, \boldu) = C_2^i(\boldx, \boldu) + \sum_{j \ne i} \big[C_3^{ij}(\boldx, \boldu) + C_4^{ij}(\boldx, \boldu) \big]$, with $Q = \diag\{2I_3, O_{3 \times 3}, 2 I_6\}$, $R = O$, and $Q_f = \diag\{10 I_3, O_{3 \times 3}, 10 I_6\}$ for each quadcopter and $Q = 2I_4$,
$R = O$, and $Q_f = O$ for each ground robot.
Using Alg. \ref{Alg: IBR} with $\alpha = 0.6$, we computed 10 trajectory rollouts (avg. runtime: 323s), which all robustly satisfied constraints (Fig. \ref{fig:heter_team_sim}).
\subsection{Heterogeneous Team Hardware Experiment}
\label{subsec: Heterogeneous Team Hardware Experiment}

\looseness-1We present a hardware analog of the heterogeneous team simulations in Sec. \ref{subsec: Heterogeneous Team Simulation}, over $T = 30$ time steps with $\Delta t = 0.4s$. 
% d
% In the hardware experiment, 
Each ground robot and quadcopter has radius $r^i = 0.06$ and $0.1$, respectively.
% wherein two robot teams, each consisting of a 3D single-integrator (i.e., waypoint-tracking) quadcopter and two 3D unicycle ground robots, interact in a shared environment.
We set $E_t^i(x_t^i) = 0.04 I_3$ for each quadcopter and 
% $E_t^i(x_t^i) = 0.001 I_4$ 
$0.001 I_4$ 
for each ground robot.
Each pair of robots must satisfy collision-avoidance constraints. The Dubins car behind the quadcopter must also satisfy proximity constraints with $\rho^{ij} = 0.75$m, and line-of-sight constraints with $\theta^{ij} = \pi/2$ rad, with respect to the quadcopter. 
% Meanwhile, the 
The other Dubins car must satisfy proximity constraints with $\rho^{ij} = 0.5$m, and line-of-sight constraints with $\theta^{ij} = \pi/3$ rad, both with respect to the first Dubins car.
When implementing Alg. 1, we set $\alpha = 0.2$, and use the cost $J^i(\boldx, \boldu) = C_2^i(\boldx, \boldu)$, with $Q = 2 I_3$, $R = O$, and $Q_f = 10 I_3$ for each quadcopter and $Q = 2 I_3$, $R = O$, and $Q_f = O$ for each ground robot.
Using Alg. \ref{Alg: IBR} with $\alpha = 0.2$, we generated robustly constraint-satisfying trajectory rollouts with an Alg. \ref{Alg: IBR} runtime of 38.3s (Fig. \ref{fig:heter_team_hardware}). 
% Our results indicate that 
Thus, our methods enable the safe, real-world deployment of heterogeneous robot teams despite uncertain dynamics.

% Here we present one simulated and one hardware experiment that involves teams composed of quadcopter and unicycle. There are 2 teams where each team has a quadcopter as leader of the first unicycle and the first unicycle acts as the leader of the second unicycle. Only quadcopters know goal positions and unicycles just follow their leaders under proximity and line-of-sight constraints. In both experiments, we add \eqref{Eqn: Coupled proximity cost} to the default trajectory cost function. 

% In the simulation, we use 12D quadcopter model and 4D unicycle model and plan for 80 time steps (0.1s per time step). The quadcopters have radius 0.05m. The unicycles have radius 0.06m and they need to ensure their leaders are within 0.5m and a 90-degree FOV. 

% In hardware experiment, we model the quadcopter model as single integrator and the ground robot as 3D Dubin's car due to the controller provided. We inflate the radius of quadcopters to be 0.1m. The proximity and line-of-sight constraints for followers of quadcopters are 0.75m and 90 degrees while 0.5m and 60 degrees for follower of unicycle. The result is shown in Fig. \ref{fig:heter_team_hardware}. 

% The simulation result in Fig. \ref{fig:heter_team_sim} illustrates that our method can design safe trajectories for multi-agent systems with high-dimensional dynamics and complex constraints, such as proximity and line-of-sight constraints. The hardware experiment in Fig. \ref{fig:heter_team_hardware} illustrates that the controller synthesized by our method ensures that each agent lies in the error bounds computed by our method. Thus, the multi-agent team satisfy all constraints robustly during real-world deployment, despite worst-case dynamics noise realizations. 

% \vspace{-9pt}
\section{Conclusion}
\label{sec: Conclusion}

\looseness-1
We present a system level synthesis (SLS)-based formulation for robust dynamic games with nonlinear dynamics, nonlinear constraints, and state-dependent disturbances. By coupling nominal planning with causal affine error feedback,
%the disturbance-to-error map is captured by 
% block lower-triangular 
%system responses,
% $\boldPhi$, 
we derive tractable nonlinear safety certificates for individual and shared constraints. We 
% build on this to 
then
introduce the robust constrained Nash equilibrium (RCNE) concept and propose an iterative best-response (IBR) method that leverages Fast SLS for scalable joint trajectory–controller optimization. In experiments, our method improved safety, proactively avoided high-disturbance regions, reduced terminal error versus baseline methods, and scaled to large, heterogeneous teams.
\revision{Our future work will develop convergence guarantees for Alg. \ref{Alg: IBR} and compare extensively against a range of single-agent and multi-agent robust motion planning baselines.
}

\vspace{-3pt}
% \printbibliography
\bibliographystyle{IEEEtran}
\bibliography{references}

% \clearpage
\appendix

% \subsection{System Level Synthesis (SLS) Preliminaries}
% \subsection{Robust Constraint Satisfaction Proof}
% \label{subsec: App, System Level Synthesis (SLS) Preliminaries}

\subsection{Proof of Proposition \ref{Prop: Robust Constraint Satisfaction}}
\label{subsec: App, Proof of Proposition on Robust Constraint Satisfaction}

We begin by writing each component of $d_t^i$ defined in \eqref{Eqn: d ti, def}, denoted $\{d_{t,k}^i: k \in [n]\}$ below, as:
\begin{align} \nonumber
    d_{t,k}^i &:= r_{t,k}^i(x_t^i, u_t^i, z_t^i, v_t^i) + E_{t,k}^i(x_t^i) w_t^i \\ \nonumber \label{Eqn: d tki, def}
    &= r_{t,k}^i(x_t^i, u_t^i, z_t^i, v_t^i) + E_{t,k}^i(z_t^i) w_t^i \\ 
    &\hspace{5mm}+ \big[E_{t,k}^i(x_t^i) - E_{t,k}^i(z_t^i) \big] w_t^i.
\end{align}

% \shuyu{Is it $E_{t,k}^i$ or $E_t^i$? }
% \frank{It is $E_{t,k}^i$; the outputs here must be scalar. }
% \begin{align}
%     % 
% \end{align}
 
We can upper bound $d_t^i$ by applying the Lagrange Theorem, similar to the process in \cite[Sec. III.B]{Leeman2025RobustNonlinearOptimalControlviaSLS}. First:
% The first and third term in \eqref{Eqn: d tki, def} can be bounded as follows:
\begin{align} \nonumber
    &|r_{t,k}^i(x_t^i, u_t^i, z_t^i, v_t^i) + \big[ E_{t,k}^i(x_t^i) -E_{t,k}^i(z_t^i) \big] w_t^i| \\
    \leq \ &\mu_{t,k}^i \Vert e_t^i \Vert_\infty^2 + L_{E,t,k}^i \Vert e_t^i \Vert_\infty.
    % = \ &\mu_{t,k}^i \Vert e_t^i \Vert_\infty^2 + L_{E,t,k}^i \Vert e_t^i \Vert_\infty^2.
\end{align}
% \shuyu{I think i\taus $L_t^i||\Delta x_t^i||_\infty$} \frank{I will refine the key assumption \#4 above to make sure the notation $L_{E,t,k}^i$ appears consistently, if that is your concern.}
Concatenating across indices $k \in [n_i]$, and introducing $\mu_t^i := \diag\{ \mu_{t,1}^i, \cdots, \mu_{t,n_i}^i \} \in \R^{n_i \times n_i}$ 
% \shuyu{I think $\mu_t^i \in R^{(n_i+m_i)\times (n_i+m_i)}$}
% \frank{Each $\mu_{t,k}^i$, for $k \in [n_i]$, is meant to be scalar, and there are a total of $n_i$ of them stacked along the diagonal of a matrix, so the overall matrix is $n_i \times n_i$.}
and $L_{E,t}^i := \diag\{L_{E,t,1}^i, \cdots, L_{E,t,n_i}^i\} \in \R^{n_i \times n_i}$, we obtain:
\begin{align} \nonumber
    &r_{t,k}^i(x_t^i, u_t^i, z_t^i, v_t^i) + \big[ E_{t,k}^i(x_t^i) -E_{t,k}^i(z_t^i) \big] w_t^i \\ \nonumber
    \in \ &\big( \Vert e_t^i \Vert_\infty^2 \mu_t^i + \Vert e_t^i \Vert_\infty L_{E,t}^i \big) \Ball_\infty^{n_i}.
\end{align}
Then:
\begin{align} \nonumber
    d_t^i &= E_t^i(z_t^i) w_t^i \\ \nonumber
    &\hspace{7.5mm} + r_{t,k}^i(x_t^i, u_t^i, z_t^i, v_t^i) + \big[ E_{t,k}^i(x_t^i) -E_{t,k}^i(z_t^i) \big] w_t^i \\ \nonumber
    &\in E_t^i(z_t^i) \Ball_\infty^{n_w} + \big( \Vert e_t^i \Vert_\infty^2 \mu_t^i + \Vert e_t^i \Vert_\infty L_{E,t}^i \big) \Ball_\infty^{n_i} \\ 
    \label{Eqn: App, d ti bounds}
    &= \begin{bmatrix}
        E_t^i(z_t^i) & \Vert e_t^i \Vert_\infty^2 \mu_t^i + \Vert e_t^i \Vert_\infty L_{E,t}^i
    \end{bmatrix} \Ball_\infty^{2n_i}.
\end{align}
% For ease of notation, we define:
% \begin{align}
%     % \label{Eqn: Lambda ti}
%     \Lambda_t^i(z) &:= \begin{bmatrix}
%         E_t^i(z_t^i) & \Vert e_t^i \Vert_\infty^2 \mu_t^i + \Vert e_t^i \Vert_\infty L_{E,t}^i
%     \end{bmatrix}.
% \end{align}
Then, recalling the definition of $\Lambda_t^i(\boldz, \boldv)$ in \eqref{Eqn: Lambda ti}, we have:
\begin{align} \label{Eqn: d ti bound, using infinity norm ball}
    d_t^i \in \Lambda_t^i(z) \ \Ball_\infty^{2n_i}
\end{align}

Now, suppose Agent $i$ applies a state feedback controller characterized by a (causal) state feedback matrix $\K^i \in \R^{(n_i + m_i)T}$ and its corresponding system response $\boldPhi^i \in \R^{(n_i + m_i)T \times n_i T}$. 
% Then, by applying SLS theory to \eqref{Eqn: Error dynamics, Delta x ti}, we obtain that:Eqn: 
% \begin{align} \label{Eqn: e from Phi and d, overall}
%     &\bolde^i := \boldPhi^i d, \\ \label{Eqn: e from Phi and d, at t}
%     \Leftrightarrow \ &e_t^i := \sum_{\tau=0}^{t-1} \boldPhi_{t-1, \tau}^i d_{t-1-\tau}^i, \hspace{5mm} \forall \ t \in [T].
% \end{align}
Then, by Taylor's Theorem, there exists a convex combination $(\tilde x_t^i, \tilde u_t^i)$ of $(x_t^i, u_t^i)$ and $(z_t^i, v_t^i)$ such that:
\begin{align} \nonumber 
% \label{Eqn: g, Taylor expansion}
    g_{t,k}^i(x_t^i, u_t^i) &= g_{t,k}^i(z_t^i, v_t^i) + \nabla g_{t,k}^i(z_t^i, v_t^i)^\top e_t^i \\ \nonumber
    &\hspace{1cm} + \frac{1}{2} (e_t^i)^\top \nabla^2 g_{t,k}^i(\tilde x_t^i, \tilde v_t^i) e_t^i.
\end{align}
% Substituting in \eqref{Eqn: e from Phi and d, at t},
Substituting $e_t^i$ with \eqref{Eqn: Disturbance Propagation, e from Phi and d},
% into \eqref{Eqn: g, Taylor expansion}, 
we obtain:
\begin{align} \nonumber
    &g_{t,k}^i(x_t^i, u_t^i) \\
    = \ &g_{t,k}^i(z_t^i, v_t^i) + \nabla g_{t,k}^i(z_t^i, v_t^i)^\top \Bigg( \sum_{\tau=0}^{t-1} \boldPhi_{t-1,\tau}^i d_{t-1, \tau}^i \Bigg) \\ \nonumber
    &\hspace{5mm} + \frac{1}{2} (e_t^i)^\top \nabla^2 g_{t,k}^i(\tilde x_t^i, \tilde v_t^i) e_t^i \\ \nonumber
    = \ &g_{t,k}^i(z_t^i, v_t^i) + \sum_{\tau=0}^{t-1} \nabla g_{t,k}^i(z_t^i, v_t^i)^\top \boldPhi_{t-1,\tau}^i d_{t-1, \tau}^i \\ 
    &\hspace{5mm} + \frac{1}{2} (e_t^i)^\top \nabla^2 g_{t,k}^i(\tilde x_t^i, \tilde v_t^i) e_t^i 
\end{align}
Further substituting in \eqref{Eqn: d ti bound, using infinity norm ball} and \eqref{Eqn: h tk Hessian bound, denoted psi tki}, we obtain:
\begin{align} \nonumber 
    &g_{t,k}^i(x_t^i, u_t^i) \\ \label{Eqn: App, Individual constraints}
    \leq \ &g_{t,k}^i(z_t^i, v_t^i)  + \chi_{t,k}^i \Vert e_t^i \Vert_\infty^2 \\ \nonumber
    &\hspace{5mm} + \sum_{\tau=0}^{t-1} \big\Vert \nabla g_{t,k}^i(z_t^i, v_t^i)^\top \boldPhi_{t-1,\tau}^i \Lambda_{t-1-\tau}(z_{t-1-\tau}^i) \big\Vert_1 
\end{align}

Similarly, for the shared constraints $h_{t,k}: \R^{n+m} \ra \R$ for each $t \in [T]$, $k \in [n_h]$, 
% , we first define:
% \begin{align} 
%     % \label{Eqn: Gamma t tau}
%     \nonumber
%     & \Gamma_{t,\tau}(\boldz, \boldv, \boldPhi) :=  \\ \nonumber
%     & \diag \big\{ \boldPhi_{t,\tau}^1 \begin{bmatrix}
%         E_{t-\tau}^1(z_{t-\tau}^1) & \Vert e_{t-\tau}^1 \Vert_\infty^2 \mu_{t-\tau}^1 + \Vert e_{t-\tau}^1 \Vert_\infty L_{E,t-\tau}^1
%     \end{bmatrix}, \cdots, \\ 
%     &\hspace{1.1cm} \boldPhi_{t,\tau}^N \begin{bmatrix}
%         E_{t-\tau}^N(z_{t-\tau}^N) & \Vert e_{t-\tau}^N \Vert_\infty^2 \mu_{t-\tau}^N + \Vert e_{t-\tau}^N \Vert_\infty L_{E,t-\tau}^N
%     \end{bmatrix} \big\}.
% \end{align}
% We then have:
we have 
\begin{align} \nonumber
    &h_{t,k}(x_t, u_t) \\ \label{Eqn: App, Shared constraints}
    \leq \ &h_{t,k}(z_t, v_t) + \psi_{t,k} \cdot \sum_{j=1}^N \Vert e_t^j \Vert_\infty^2 \\ \nonumber
    &\hspace{5mm} + \sum_{\tau=0}^{t-1} \big\Vert \nabla h_{t,k}(z_t, v_t)^\top \Gamma_{t-1,\tau}(\boldz, \boldv, \boldPhi).
\end{align}

% Finally, we introduce the auxiliary error bounds $\boldrho^i := \{\rho_t^i \}$ that appear in \eqref{Eqn: rho initial bound at t = 0, to use in opt} and \eqref{Eqn: rho recursive bound, to use in opt} in Prop. \ref{Prop: Robust Constraint Satisfaction}. Moreover, for each $i \in [N]$, we establish inductive bounds on $\rho_t^i$ inductively across times $t \in [T]$ to form \eqref{Eqn: rho recursive bound, to use in opt}, 

Finally, we show 
% using \eqref{Eqn: App, Individual constraints} and \eqref{Eqn: App, Shared constraints} 
that by leveraging the bounds \eqref{Eqn: rho initial bound at t = 0, to use in opt} and \eqref{Eqn: rho recursive bound, to use in opt} in Prop. \ref{Prop: Robust Constraint Satisfaction}, repeated below, we can use \eqref{Eqn: App, Individual constraints} and \eqref{Eqn: App, Shared constraints} to establish \eqref{Eqn: g tki upper bound, to use in opt} and \eqref{Eqn: h tk upper bound, to use in opt} 
% (as reiterated below) 
respectively:
\begin{align} \nonumber
    &\rho_0^i \geq 0, \\ \nonumber
    &\sum_{\tau=0}^{t-1} \Vert \boldPhi_{t-1, \tau}^i \begin{bmatrix}
        E_\tau^i(z_\tau^i) & (\rho_\tau^i)^2 \mu_\tau^i + \rho_\tau^i L_{E,\tau}^i
    \end{bmatrix} \Vert_\infty \leq \rho_t^i.
\end{align}
Our key step is to show, via induction, that \eqref{Eqn: rho initial bound at t = 0, to use in opt} and \eqref{Eqn: rho recursive bound, to use in opt} imply $\Vert e_t^i \Vert_\infty \leq \rho_t^i$ for each $i \in [N]$, $t \in [0, T]$. To this end, fix $i \in [N]$, and consider the induction hypothesis that, for some $t \in [0, T]$, we have $\Vert e_\tau^i \Vert_\infty \leq \rho_\tau^i$ for all $\tau \leq t-1$. For $t = 0$, this induction hypothesis is vacuously true; for $t = 1$, it is true via \eqref{Eqn: rho initial bound at t = 0, to use in opt}. Then, from the inclusion relation \eqref{Eqn: App, d ti bounds} associated with $d_t^i$, we have:
\begin{align}
    d_t^i &\in \begin{bmatrix}
        E_t^i(z_t^i) & \Vert e_t^i \Vert_\infty^2 \mu_t^i + \Vert e_t^i \Vert_\infty L_{E,t}^i 
    \end{bmatrix} \Ball_\infty^{2n_i} \\
    &\subseteq \begin{bmatrix}
        E_t^i(z_t^i) & (\rho_t^i)^2 \mu_t^i + \rho_t^i L_{E,t}^i
    \end{bmatrix} \Ball_\infty^{2n_i}.
\end{align}
Then:
\begin{align} \nonumber 
    \Vert e_t^i \Vert_\infty &= \left\Vert \sum_{\tau=0}^{t-1} \boldPhi_{t-1,\tau}^i d_\tau^i \right\Vert_\infty \\ \nonumber
    &\leq \sum_{\tau=0}^{t-1} \big\Vert \boldPhi_{t-1,\tau}^i \begin{bmatrix}
        E_\tau^i(z_\tau^i) & (\rho_\tau^i)^2 \mu_t^i + \rho_\tau^i L_{E,\tau}^i
    \end{bmatrix} \big\Vert_\infty \\ \label{Eqn: tube_error_overbound}
    &\leq \rho_t^i.
\end{align}
We then substitute \eqref{Eqn: tube_error_overbound} into \eqref{Eqn: App, Individual constraints} and \eqref{Eqn: App, Shared constraints} to obtain the constraints \eqref{Eqn: g tki upper bound, to use in opt} and \eqref{Eqn: h tk upper bound, to use in opt} in Prop. \ref{Prop: Robust Constraint Satisfaction}. 
which guarantee that the original constraints \eqref{Eqn: Individual Constraints} and \eqref{Eqn: Shared Team Constraints} are robustly satisfied despite worst-case noise realizations.

% \end{proof}

% \input{temp___Alg_poster}

% \input{A2_Experiment_Details}

%%%%%%%%%%%%%%%%%%%%%%%%%%%%%%%%%%%%%%%%%%%%%%%%%%%%%%%%%%%%%%%%%%%%%%%%%%%%%%%%
% \section*{APPENDIX}

% \input{old_1___App_A1_algorithm_details_20250915_1042.}

% \input{old_2___Antoine_algorithm_notes}

% \input{old_3___Preliminaries_20250915_1042}

% \input{old_4___Methods_20250915_1042}

% \input{old_5___Experiments}

% \end{thebibliography}

% \section*{ACKNOWLEDGMENT}

\end{document}